\documentclass[letterpaper,12pt,leqno]{article}
\usepackage{paper,math}
\pdfoutput=1
\def\bib{paper.bib}
\def\pdf{figures.pdf}
\published{Journal of Public Economics Plus}{https://www.pascalmichaillat.org/9.html}
\hypersetup{pdftitle={Beveridgean Unemployment Gap}}

\def\W{\wh{\Wc}}

\begin{document}

\author{Pascal Michaillat, Emmanuel Saez}
\title{Beveridgean Unemployment Gap
\thanks{Michaillat: Brown University. Saez: University of California--Berkeley. We thank Regis Barnichon, Olivier Blanchard, Michael Boskin, Varanya Chaubey, Raj Chetty, Gabriel Chodorow-Reich, Richard Crump, Peter Diamond, Andrew Figura, Nathaniel Hendren, Damon Jones, Marianna Kudlyak, Etienne Lehmann, Sephorah Mangin, Adam McCloskey, Benjamin Schoefer, Johannes Spinnewijn, Alison Weingarden, Roberton Williams, and Owen Zidar for helpful comments and discussions. This work was supported by the Institute for Advanced Study and the Berkeley Center for Equitable Growth.}}
\date{November 2021}

\begin{titlepage}\maketitle

This paper develops a sufficient-statistic formula for the unemployment gap---the difference between the actual unemployment rate and the efficient unemployment rate. While lowering unemployment puts more people into work, it forces firms to post more vacancies and to devote more resources to recruiting. This unemployment-vacancy tradeoff, governed by the Beveridge curve, determines the efficient unemployment rate. Accordingly, the unemployment gap can be measured from three sufficient statistics: elasticity of the Beveridge curve, social cost of unemployment, and cost of recruiting. Applying this formula to the United States, 1951--2019, we find that the efficient unemployment rate averages $4.3\%$, always remains between $3.0\%$ and $5.4\%$, and has been stable between $3.8\%$ and $4.6\%$ since 1990. As a result, the unemployment gap is countercyclical, reaching 6 percentage points in slumps. The US labor market is therefore generally inefficient and especially inefficiently slack in slumps. In turn, the unemployment gap is a crucial statistic to design labor-market and macroeconomic policies.

\end{titlepage}\section{Introduction}

\paragraph{Research question} Does the labor market operate efficiently? If not, how far from efficiency is it? To answer this question, we develop a new measure of the unemployment gap---the difference between the actual unemployment rate and the efficient unemployment rate. A reliable measure of the unemployment gap is necessary to a good understanding of the labor market and macroeconomy: it guides how we model the labor market and macroeconomy; and it is a key determinant of optimal labor-market policies, such as unemployment insurance, and of optimal macroeconomic policies, including monetary policy and fiscal policy.

\paragraph{Existing measures of the unemployment gap} Two measures of the unemployment gap are commonly used: the difference between the unemployment rate and its trend; and the difference between the unemployment rate and the non-accelerating-inflation rate of unemployment (NAIRU). Although these two measures are easy to use, neither is an ideal measure of the unemployment gap because neither the unemployment-rate trend nor the NAIRU measure the efficient unemployment rate.

\paragraph{Our measure of the unemployment gap} This paper proposes a new measure of the unemployment gap. The measure is based upon the theory of efficiency in modern labor-market models \cp{H90,P00}. Such models feature both unemployed workers and job vacancies, each inducing welfare costs: more unemployment means fewer people at work so less output; more vacancies means more labor devoted to recruiting and also less output. Furthermore, these models feature a Beveridge curve: a negative relation between unemployment and vacancies. Because of the Beveridge curve, unemployment and vacancies cannot be simultaneously reduced: less unemployment requires more vacancies, and fewer vacancies create more unemployment. Our analysis resolves this unemployment-vacancy tradeoff by characterizing the efficiency point on the Beveridge curve.

To obtain an unemployment-gap measure that is easy to use, we express the unemployment gap in terms of sufficient statistics \cp{C09}. The sufficient-statistic formula is simple, involving only three statistics. The formula is also easy to apply because the statistics are estimable. Finally, the formula is valid in a broad class of models, including in the widely used Diamond-Mortensen-Pissarides (DMP) model.

\paragraph{Unemployment-gap formula} To characterize the efficient unemployment rate, we solve the problem of a social planner who allocates labor between production, recruiting, and unemployment subject to the Beveridge curve. We then express the efficient unemployment rate as a function of three sufficient statistics: elasticity of the Beveridge curve, social cost of unemployment, and cost of recruiting. The difference between the actual unemployment rate and the efficient unemployment rate gives the unemployment gap.

\paragraph{Comparison with the \ct{H90} condition} We apply our efficiency condition to the DMP model in order to compare it with the well-known Hosios condition. The two might differ because they solve different planning problems: in the Hosios planning problem unemployment follows a differential equation, whereas in ours unemployment is always on the Beveridge curve. We find that when the discount rate is zero, the two conditions coincide. When the discount rate is positive, the two conditions differ, but they produce almost identical allocations.

\paragraph{Estimates of the sufficient statistics in the United States} Next we estimate the three sufficient statistics in the United States. Although the Beveridge curve is stable for long periods, it also experiences sudden shifts. To estimate the Beveridge elasticity in the presence of these structural breaks, we use the algorithm proposed by \ct{BP98,BP03}. We find that between 1951 and 2019, the Beveridge curve experiences $5$ structural breaks, but the Beveridge elasticity remains stable between $0.84$ and $1.02$. Next, we estimate the social cost of unemployment from the experimental evidence presented by \ct{BM15} and \ct{MP19}. We find that the value from home production and recreation during unemployment only replaces $26\%$ of the marginal product of labor---implying a substantial social cost of unemployment. Last, following \ct{V10}, we estimate the recruiting cost from the 1997 National Employer Survey. We find that firms allocate $3.2\%$ of their labor to recruiting.

\paragraph{Unemployment gap in the United States} Using the estimated statistics, we compute the efficient unemployment rate and unemployment gap in the United States from 1951 to 2019. The efficient unemployment rate averages $4.3\%$, and it always remains between $3.0\%$ and $5.4\%$. It started at $3.5\%$ in 1951, climbed to $5.4\%$ in 1978, fell to $4.6\%$ in 1990, and remained between $3.8\%$ and $4.6\%$ thereafter.

As the efficient unemployment rate remains within a narrow band while the actual unemployment rate fluctuates widely, the unemployment gap is almost never zero. Hence, the US labor market is generally inefficient. In fact, the unemployment gap is generally positive, averaging $1.4$ percentage points, so the US labor market is generally inefficiently slack. Furthermore, just like the unemployment rate, the unemployment gap is countercyclical. For instance, the unemployment gap reached $6.2$ percentage points in 2009, in the aftermath of the Great Recession. Therefore, the US labor market is especially inefficiently slack in slumps.\footnote{These findings are consistent with those in \ct[figure 3]{LMS15}. They find that the US labor market is generally inefficient, and is inefficiently slack in slumps. However, by using a simpler model without insurance considerations, we are able to obtain further results. First, we derive a formula for the unemployment gap and measure the gap, instead of just signing it. Second, our formula applies to any Beveridgean model, not just the matching model. Third, we obtain a diagrammatical representation of efficiency, which clarifies conceptual issues.}

\paragraph{Robustness} We explore the sensitivity of the US unemployment gap to alternative values of the sufficient statistics. We find that for all plausible estimates of the sufficient statistics, the unemployment gap remains within $1.2$ percentage points of its baseline. Thus our conclusions that the US labor market is generally inefficient and inefficiently slack in slumps are robust to alternative calibrations.

\section{Beveridgean model of the labor market}\label{s:model}

We introduce the model of the labor market used to compute the unemployment gap. The main feature of the model is a Beveridge curve: a negative relation between the number of unemployed workers and the number of job vacancies.

\subsection{Notations and definitions}

The unemployment rate $u \in (0,1)$ is the number of unemployed workers divided by the size of the labor force. The vacancy rate $v\in (0,\infty)$ is the number of vacancies divided by the size of the labor force. The labor-market tightness $\t = v/u$ is the number of vacancies per unemployed worker. The employment rate $n\in (0,1)$ is the number of employed workers divided by size of the labor force. Since the labor force consists of all employed and unemployed workers, employment and unemployment rates are related by $n=1-u$.

\subsection{Beveridge curve}

The only restriction imposed on the labor market is that unemployment and vacancy rates are related by a Beveridge curve:

\begin{assumption}\label{a:v} The vacancy rate is given by a differentiable, strictly decreasing, and strictly convex function of the unemployment rate, denoted $v(u)$. \end{assumption}

\begin{figure}[p]
\subcaptionbox{Unemployment rate\label{f:u}}{\includegraphics[scale=\sfig,page=1]{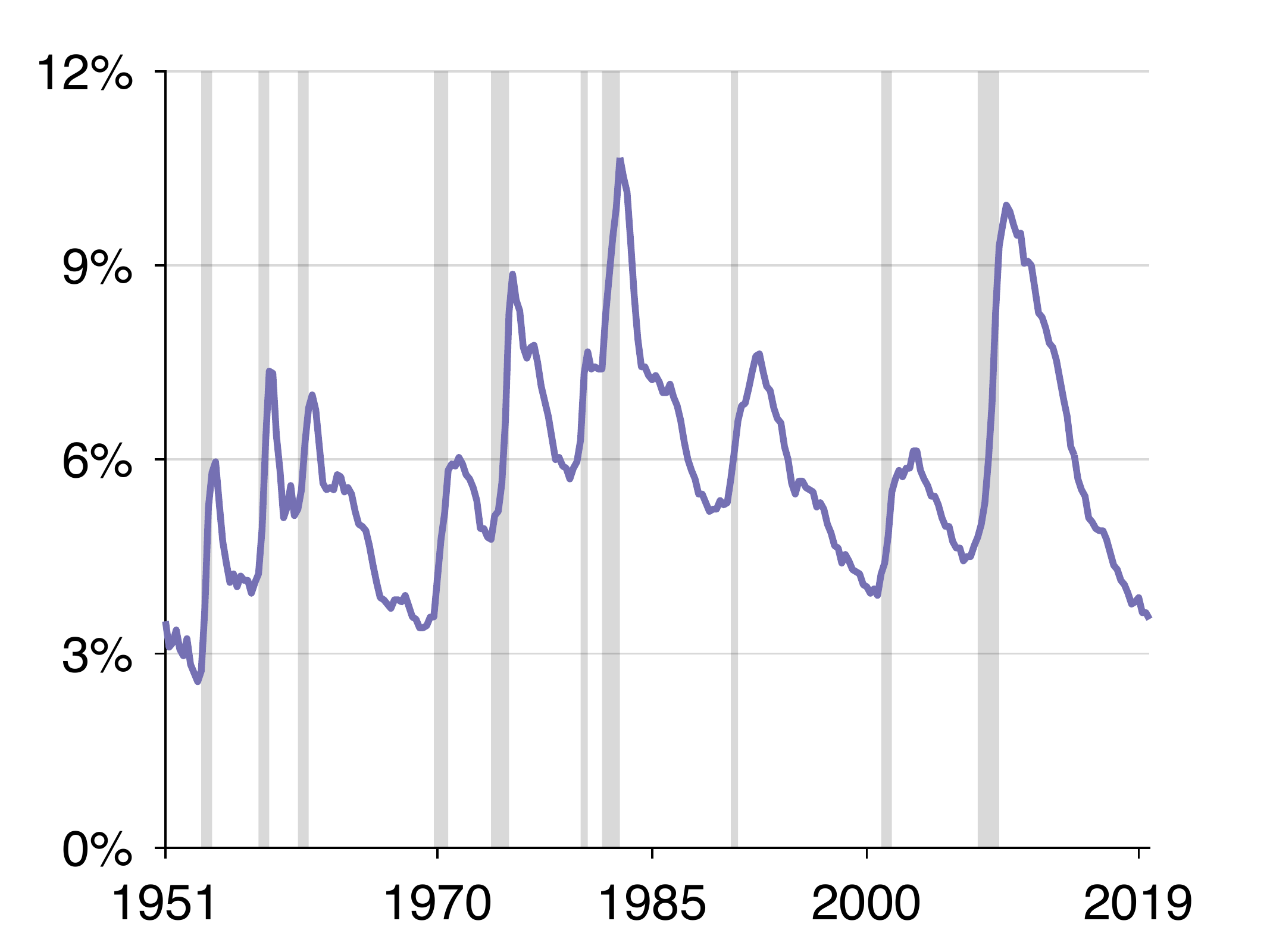}}\hfill
\subcaptionbox{Vacancy rate\label{f:v}}{\includegraphics[scale=\sfig,page=2]{\pdf}}\\
\subcaptionbox{Beveridge curve, 1951--1969\label{f:beveridge1}}{\includegraphics[scale=\sfig,page=3]{\pdf}}\hfill
\subcaptionbox{Beveridge curve, 1970--1989\label{f:beveridge2}}{\includegraphics[scale=\sfig,page=4]{\pdf}}\\
\subcaptionbox{Beveridge curve, 1990--2009\label{f:beveridge3}}{\includegraphics[scale=\sfig,page=5]{\pdf}}\hfill
\subcaptionbox{Beveridge curve, 2010--2019\label{f:beveridge4}}{\includegraphics[scale=\sfig,page=6]{\pdf}}
\caption{Beveridge curve in the United States, 1951--2019}
\note{A: The unemployment rate is constructed by the \ct{UNRATE}. B: For 1951--2000, the vacancy rate is constructed by \ct{B10d}; for 2001--2019, the vacancy rate is the number of job openings divided by the civilian labor force, both measured by the \ct{CLF16OV,JTSJOL}. Unemployment and vacancy rates are quarterly averages of monthly series. The shaded areas are NBER-dated recessions. C--F: Unemployment and vacancy rates come from panels~A and B. For readability, we separately plot the 1951--1969, 1970--1989, 1990--2009, and 2010--2019 periods.}
\label{f:beveridge}\end{figure}

\paragraph{Beveridge curve in the data} The Beveridge curve appears in many countries \cp{JPS90,BD89,NNO02,DS15,EMR15}. As an illustration, we construct the Beveridge curve in the United States, 1951--2019. We use the standard measure of unemployment rate, constructed by the \ct{UNRATE} from the Current Population Survey (figure~\ref{f:u}). We measure the vacancy rate from two different sources, because there is no continuous vacancy series over the period. For 1951--2000, we use the vacancy proxy constructed by \ct{B10d}. \name{B10d} starts from the help-wanted advertising index constructed by the Conference Board. He then corrects the Conference Board index, which is based on newspaper advertisements, to take into account the shift from print advertising to online advertising in the 1990s. Finally, he rescales the index into vacancies and and divides it by the size of the labor force to obtain a vacancy rate. For 2001--2019, we use the number of job openings measured by the \ct{JTSJOL} in the Job Opening and Labor Turnover Survey, divided by the civilian labor force constructed by the \ct{CLF16OV} from the Current Population Survey. We then splice the two series to obtain a vacancy rate for 1951--2019 (figure~\ref{f:v}).

The Beveridge curve appears in scatterplots of the unemployment and vacancy rates (figures~\ref{f:beveridge1}--\ref{f:beveridge4}). The Beveridge curve is stable for long periods, during which unemployment and vacancies move up and down along a clearly defined branch. Furthermore, until the mid-1980s, the Beveridge curve shifts outward at the end of each period of stability. After the mid-1980s, the Beveridge curve shifts back inward to positions that were typical in the 1960s and 1970s. On a logarithmic scale, each branch of the Beveridge curve is close to linear; hence, each branch is close to isoelastic. 

\paragraph{Microfoundations for the Beveridge curve} The Beveridge curve may arise from several microfoundations \cp{EMR15}. Labor-market models built around a matching function feature a Beveridge curve: for instance, the DMP model \cp{P00,S05} and its variants with rigid wages \cp{H05,HM08}, large firms \cp{CMW08,EM08}, and job rationing \cp{M09}. Macroeconomic models in which self-employed workers sell labor services to consumers on a matching market also feature a Beveridge curve \cp{MS15,MS19}. Other microfoundations for the Beveridge curve include mismatch \cp{S07} and stock-flow matching \cp{ES10}.

It is true that in many matching models, unemployment follows a law of motion, and the Beveridge curve is defined as the locus of unemployment and vacancies where unemployment remains steady. Yet \ct[p.~236]{Pi09} notes that
\begin{quote}
``Perhaps surprisingly at first, but on reflection not so surprisingly, we get a good approximation to the dynamics of unemployment if we treat unemployment as if it were always on the Beveridge curve.''
\end{quote}
Hence, many matching models assume that the Beveridge curve holds at all times, as we do here \cp{P86,P07,H05,Ha05,EMS09}.

\paragraph{Fluctuations along the Beveridge curve} Over the business cycle, unemployment and vacancy rates move along the Beveridge curve (figure~\ref{f:beveridge}). What causes such fluctuations? In the DMP model, shocks to workers' bargaining power lead to fluctuations along the Beveridge curve \cp[table~6]{S05}. Shocks to labor productivity also lead to fluctuations along the Beveridge curve, but these are much smaller than empirical fluctuations \cp[table~3]{S05}. In the variants of the DMP model by \ct{H05}, \ct{HM08}, and \ct{M09}, real wages are rigid, and shocks to labor productivity generate realistic fluctuations along the Beveridge curve. In the mismatch model by \ct{S07} and stock-flow matching model by \ct{ES10}, shocks to labor productivity also generate sizable fluctuations along the Beveridge curve. Finally, in the macroeconomic models by \ct{MS15,MS19}, aggregate-demand shocks generate fluctuations along the Beveridge curve.

\subsection{Beveridge elasticity} 

The Beveridge curve governs the tradeoff between unemployment and vacancies. The elasticity of the Beveridge curve therefore plays a central role in the analysis.

\begin{definition} The Beveridge elasticity is the elasticity of the vacancy rate with respect to the unemployment rate along the Beveridge curve, normalized to be positive: 
\begin{equation}
\e = -\odl{v(u)}{u} = - \frac{u}{v} \cdot v'(u).
\label{e:elasticity}\end{equation}\end{definition}

\subsection{Social welfare}

The unemployment-vacancy tradeoff is a the core of our analysis because both unemployment and vacancies induce welfare costs. Consider for instance the DMP model \cp{P00}. The labor force is composed of $L>0$ workers with linear utility function. Employed workers have a productivity $p>0$. Unemployed workers engage in home production but they are less productive than employed workers; their productivity is $p\cdot z <p$.\footnote{While \ct[pp.~13, 21, 72, 74]{P00} initially specifies the productivity of unemployed workers as constant---independent of the productivity of employed workers---he subsequently considers specifications in which the productivity of unemployed workers is proportional to that of employed workers. We model the productivity of unemployed workers as proportional to that of employed workers to be consistent with the evidence presented by \ct{CK16}.} Firms incur a flow recruiting cost $p\cdot c > 0$ for each vacancy. Hence, flow social welfare is given by the linear function
\begin{equation}
\Wc(n,u,v) = p \bp{n  + z u - c v} L.
\label{e:w}\end{equation}

However, it is possible to be more general:
 
\begin{assumption}\label{a:w} Flow social welfare is a function of the employment rate, unemployment rate, and vacancy rate, denoted $\Wc(n,u,v)$. The function $\Wc$ is differentiable, strictly increasing in $n$, strictly decreasing in $v$, and less increasing in $u$ than $n$ (so $\pdx{\Wc}{u}<\pdx{\Wc}{n}$). As a result, the alternate welfare function $\W(u,v) = \Wc(1-u,u,v)$ is strictly decreasing in $u$ and $v$. Furthermore, the function $\W$ is quasiconcave.\end{assumption}

Employed workers contribute to social welfare through market production, which is why $\pdx{\Wc}{n}>0$. Unemployed workers contribute to social welfare through home production and recreation \cp{AHK13}; this contribution is diminished if people suffer psychic pain from being unemployed \cp{Br15,HKL21}. However, unemployed workers contribute less to welfare than employed workers, so $\pdx{\Wc}{u}<\pdx{\Wc}{n}$. Vacancies reduce social welfare because to fill a vacancy, labor and other resources must be diverted from market production toward recruiting. 

The alternate welfare function $\W(u,v)$ is obtained from the welfare function $\Wc(n,u,v)$ by substituting the employment rate $n$ by $1-u$. The property that the alternate welfare function decreases with the unemployment and vacancy rates captures the welfare costs of unemployment and vacancies. We assume that the alternate welfare function is quasiconcave to ensure that the social planner's problem is well behaved.

\subsection{Social value of nonwork} 

To measure the welfare cost induced by unemployment, we introduce the following statistic:

\begin{definition} The social value of nonwork is the marginal rate of substitution between unemployment and employment in the welfare function:
\begin{equation*}
\z = \frac{\pdx{\Wc}{u}}{\pdx{\Wc}{n}}<1.
\end{equation*}
The social cost of unemployment is 
\begin{equation*}
\frac{(\pdx{\Wc}{n})-(\pdx{\Wc}{u})}{\pdx{\Wc}{n}} = 1 - \z>0.
\end{equation*}
\end{definition}

The social value of nonwork $\z$ measures the marginal contribution of unemployed workers to welfare, relative to that of employed workers. It is less than $1$ because unemployed workers' contribute less to welfare than employed workers (assumption~\ref{a:w}). The social cost of unemployment $1-\z>0$ measures the welfare loss from having a person unemployed rather than employed. Such loss comprises foregone market production and the psychological pain of being unemployed, net of home production and the value of recreation when unemployed.

\subsection{Recruiting cost} 

To measure the welfare cost induced by vacancies, we introduce the following statistic:

\begin{definition} The recruiting cost is the marginal rate of substitution between vacancies and employment in the welfare function, normalized to be positive:
\begin{equation*}
\k = -\frac{\pdx{\Wc}{v}}{\pdx{\Wc}{n}} > 0.
\end{equation*}\end{definition}

The recruiting cost $\k$ measures the number of recruiters allocated to each vacancy. These workers are tasked with writing and advertising the vacancy; reading applications and finding suitable candidates; interviewing and evaluating selected candidates; and drafting and negotiating job offers.

\section{Beveridgean unemployment gap}

We solve the problem of a social planner who chooses the unemployment and vacancy rates to maximize welfare subject to the Beveridge curve. The solution gives the efficient unemployment rate. To understand the tradeoffs at play, we represent efficiency in a Beveridge diagram. To be able to measure the efficient unemployment rate empirically, we express it with sufficient statistics. The difference between the actual unemployment rate and the efficient unemployment rate then gives the unemployment gap.

\subsection{Social planner's problem}

To find the efficient unemployment rate, we solve the problem of a social planner who is subject to the Beveridge curve:

\begin{definition} The efficient unemployment and vacancy rates, denoted $u^*$ and $v^*$, maximize social welfare $\W(u,v)$ subject to the Beveridge curve $v=v(u)$. The efficient labor-market tightness is $\t^* = u^* / v^*$, and the unemployment gap is $u-u^*$.\end{definition}

The social planner's problem generalizes the problem in \ct{H90}, in that it covers any model with a Beveridge curve, not just those with a matching function, and any quasiconcave welfare function, not just linear ones.

\paragraph{Comparison of the planning and decentralized solutions} The planning solution is described by two variables: unemployment and vacancies. These variables are given by two equations: the Beveridge curve, and the first-order condition of the planner's problem. By contrast, the decentralized solution is usually given by three variables: unemployment, vacancies, and wages. These variables are usually given by three equations: the Beveridge curve; a wage equation; and an equation describing vacancy creation, such as the free-entry condition in the DMP model. 

In many Beveridgean models, unlike in Walrasian models, the decentralized solution does not overlap with the planning solution. This is because most wage mechanisms do not ensure efficiency. For instance, in matching models, the wage is determined in a situation of bilateral monopoly, so a range of wages is theoretically possible. A wage mechanism picks one wage within the range. There is only an infinitesimal chance that this wage is the one ensuring efficiency \cp[p.~185]{P00}. Accordingly, theory does not guarantee that the unemployment gap is zero.

\subsection{Efficiency in the Beveridge diagram}

To illustrate the tradeoffs facing the social planner, we represent labor-market efficiency in a Beveridge diagram (figure~\ref{f:theory}).

\paragraph{Beveridge diagram} The Beveridge diagram features unemployment rate on the $x$-axis and vacancy rate on the $y$-axis. The diagram displays the Beveridge curve: the locus of unemployment and vacancy rates that are feasible in the economy. The Beveridge curve is downward-sloping and convex. The diagram also displays an isowelfare curve: the locus of unemployment and vacancy rates such that social welfare $\W(u,v)$ remains constant at some level. Since $\W(u,v)$ is decreasing in both arguments, the points inside the isowelfare curve yield higher welfare, so the green area is an upper contour set of $\W(u,v)$. Since the function $\W$ is quasiconcave, the upper contour sets are convex.

\begin{figure}[t]
\subcaptionbox{Efficient unemployment rate\label{f:efficiency}}{\includegraphics[scale=\sfig,page=7]{\pdf}}\hfill
\subcaptionbox{Unemployment gap\label{f:inefficiency}}{\includegraphics[scale=\sfig,page=8]{\pdf}}
\caption{Efficient unemployment rate and unemployment gap}
\note{The Beveridge curve has slope $v'(u)$. The isowelfare curve has slope $-(1-\z)/\k$, where $\z$ is the social value of nonwork and $\k$ is the recruiting cost. The tangency of the Beveridge and isowelfare curves gives the efficient labor-market allocation. Other allocations along the Beveridge curve are inefficient and feature a nonzero unemployment gap.}
\label{f:theory}\end{figure}

\paragraph{Efficiency condition} The efficient unemployment and vacancy rates can easily be found in the Beveridge diagram (figure~\ref{f:efficiency}). First, they have to lie on the Beveridge curve. Second, since both unemployment and vacancies reduce welfare, they must lie on an isowelfare curve that is as close as possible to the origin. The closest the isowelfare curve can be while maintaining contact with the Beveridge curve is at their tangency point. Hence, the efficient unemployment and vacancy rates are found at the point where the Beveridge curve is tangent to an isowelfare curve.

Furthermore, the slope of the Beveridge curve is $v'(u)$. The slope of the isowelfare curve is minus the marginal rate of substitution between unemployment and vacancies in the welfare function $\W(u,v)$:
\begin{equation*}
-\frac{\pdx{\W}{u}}{\pdx{\W}{v}} = -\frac{(\pdx{\Wc}{u})-(\pdx{\Wc}{n})}{\pdx{\Wc}{v}} = -\frac{1-(\pdx{\Wc}{u})/(\pdx{\Wc}{n})}{-(\pdx{\Wc}{v})/(\pdx{\Wc}{n})} = -\frac{1-\z}{\k}.
\end{equation*}

We infer the following result:

\begin{proposition}\label{p:plan} In a Beveridge diagram, efficiency is achieved at the point where the Beveridge curve is tangent to an isowelfare curve. Hence, the efficient unemployment rate is implicitly defined by 
\begin{equation}
v'(u)  = - \frac{1-\z}{\k},
\label{e:tangent}\end{equation}
where $\z<1$ is the social value of nonwork and $\k>0$ is the recruiting cost.\end{proposition}

Formula \eqref{e:tangent} says that when the labor market operates efficiently, welfare costs and benefits from moving one worker from employment to unemployment are equalized. Moving one worker from employment to unemployment reduces welfare by the social cost of unemployment, $1-\z$. Having one more unemployed worker also means having $-v'(u)>0$ fewer vacancies. Each vacancy reduces welfare by the recruiting cost, $\k$, so welfare improves by $-v'(u) \k$. When welfare costs and benefits are equalized, we have $1-\z = - v'(u) \k$, which is equivalent to \eqref{e:tangent}.

\paragraph{Deviations from efficiency} The labor market may not operate efficiently (figure~\ref{f:inefficiency}). The labor market may be above the efficiency point, where unemployment is too low, vacancies are too high, tightness is too high, and the unemployment gap is negative. It may also be below the efficiency point, where unemployment is too high, vacancies are too low, tightness is too low, and the unemployment gap is positive. As both allocations are inefficient, they lie on a worse isowelfare curve than the efficiency point.

\begin{figure}[t!]
\subcaptionbox{Compensated increase in Beveridge elasticity\label{f:comparativeEpsilon}}{\includegraphics[scale=\sfig,page=11]{\pdf}}\hfill
\subcaptionbox{Increase in social value of nonwork\label{f:comparativeZeta}}{\includegraphics[scale=\sfig,page=9]{\pdf}}\vfig
\subcaptionbox{Increase in recruiting cost\label{f:comparativeKappa}}{\includegraphics[scale=\sfig,page=10]{\pdf}}
\caption{Comparative statics for the efficient unemployment rate}
\note{Each panel reproduces figure~\ref{f:efficiency} before and after an increase in one of the statistics. A: The efficient unemployment rate increases when the Beveridge elasticity increases, keeping welfare constant. B: The efficient unemployment rate increases when the social value of nonwork increases. C: The efficient unemployment rate increases when the recruiting cost increases.}
\label{f:comparative}\end{figure}

\paragraph{Comparative statics} We use the Beveridge diagram to derive several comparative statics (figure~\ref{f:comparative}). We first consider a compensated increase in the Beveridge elasticity (analogous to a compensated price increase in the context of Hicksian demand). This is an increase in the Beveridge elasticity, $\e$, compensated by a shift of the Beveridge curve so that it remains tangent to the same isowelfare curve. Such increase steepens the Beveridge curve (figure~\ref{f:comparativeEpsilon}). At the previous efficiency point, the Beveridge curve is steeper than the isowelfare curve. Therefore, the new efficiency point must be lower than the old one on the isowelfare curve, and the Beveridge curve must shift out to maintain welfare at the same level. Thus, the efficient unemployment rate is higher than before. The intuition is simple: when a rise in unemployment triggers a larger drop in vacancies, the unemployment-vacancy tradeoff is more favorable to unemployment.

Next, we consider an increase in the social value of nonwork, $\z$. The isowelfare curve has slope $-(1-\z)/\k$, so it becomes flatter everywhere (figure~\ref{f:comparativeZeta}). At the previous efficiency point the isowelfare curve is flatter than the Beveridge curve, so the new efficiency point is lower on the Beveridge curve. Hence, the efficient unemployment rate is higher than before. Intuitively, when unemployment is less costly, the unemployment-vacancy tradeoff is more favorable to unemployment.

Finally, we consider an increase in recruiting cost, $\k$. The isowelfare curve is flatter everywhere, so the efficient unemployment rate is higher (figure~\ref{f:comparativeKappa}). Intuitively, when recruiting is more costly, the unemployment-vacancy tradeoff is more favorable to unemployment. 

The following corollary summarizes the comparative statics:

\begin{corollary} The efficient unemployment rate increases after a compensated increase in the Beveridge elasticity (increase in elasticity keeping welfare constant), after an increase in the social value of nonwork, and after an increase in the recruiting cost.\end{corollary}

\subsection{Efficient labor-market tightness}

Toward obtaining a sufficient-statistic formula for the unemployment gap, we rework efficiency condition \eqref{e:tangent} and derive a sufficient-statistic formula for the efficient tightness. We begin by introducing the Beveridge elasticity \eqref{e:elasticity}, which satisfies $\e\t = - v'(u)$. With this expression, we rewrite \eqref{e:tangent} as $\t = (1-\z)/(\k\e)$. In figure~\ref{f:inefficiency}, we see that any point on the Beveridge curve above the efficiency point has $- v'(u) > (1-\z)/\k$, and any point below it has $- v'(u) < (1-\z)/\k$. Using again $\e\t = - v'(u)$, we infer that any point above the efficiency point satisfies $\t> (1-\z)/(\k\e)$; and any point below the efficiency point has $\t < (1-\z)/(\k\e)$.

Hence we can assess the efficiency of tightness from three sufficient statistics:

\begin{proposition}\label{p:theta} Consider a point on the Beveridge curve with tightness $\t$, Beveridge elasticity $\e$, social value of nonwork $\z$, and recruiting cost $\k$. Tightness is inefficiently high if $\t > (1-\z)/(\k\e)$,  inefficiently low if $\t < (1-\z)/(\k\e)$, and efficient if 
\begin{equation}
\t = \frac{1-\z}{\k \e}.
\label{e:theta}\end{equation}\end{proposition}

Since the statistics $\e$, $\z$, and $\k$ generally depend on tightness $\t$, formula~\eqref{e:theta} only characterizes the efficient tightness implicitly. This limitation is typical of the sufficient-statistic approach \cp{C09}. 

\subsection{Efficient unemployment rate and unemployment gap}

To compute the efficient unemployment rate and unemployment gap, we must address the endogeneity of the sufficient statistics in \eqref{e:theta}. We use a workaround developed by \ct{K18}:

\begin{assumption}\label{a:constant} The Beveridge elasticity $\e$, social value of nonwork $\z$, and recruiting cost $\k$  do not depend on the unemployment and vacancy rates.\end{assumption}

How realistic is this assumption? Figure~\ref{f:beveridge} suggests that the Beveridge curve is isoelastic, so the assumption on the Beveridge elasticity seems valid in the United States. We do not have comparable evidence on the social value of nonwork and recruiting cost, but at least in the DMP model, these two statistics are independent of the unemployment and vacancy rates (equation \eqref{e:w}).

Under assumption~\ref{a:constant}, the Beveridge curve is isoelastic:
\begin{equation*}
 v(u)=\a \cdot u^{-\e},
\end{equation*}
where the parameter $\a>0$ determines the location of the curve. On the Beveridge curve, tightness and unemployment are related by $\t = v(u)/u = \a \cdot u^{-(1+\e)}$, and efficient tightness and unemployment are related by $\t^* = \a \cdot (u^*)^{-(1+\e)}$. Moreover, $\a$ can be computed from the vacancy and unemployment rates: $\a = v/u^{-\e}$. Hence, the efficient unemployment rate is given by 
\begin{equation*}
u^* = \bp{\frac{1}{\t^*}\cdot \frac{v}{u^{-\e}}}^{1/(1+\e)}.
\end{equation*}

Under assumption~\ref{a:constant}, formula \eqref{e:theta} also explicitly determines the efficient tightness:
\begin{equation*}
\t^* = \frac{1-\z}{\k\e}.
\end{equation*}

Combining the last two equations, we obtain the following result:

\begin{proposition}\label{p:u} Under assumption~\ref{a:constant}, the efficient unemployment rate $u^*$ can be measured from current unemployment rate $u$, vacancy rate $v$, Beveridge elasticity $\e$, social value of nonwork $\z$, and recruiting cost $\k$:
\begin{equation}
u^* = \bp{\frac{\k \e}{1-\z}\cdot\frac{v}{u^{-\e}}}^{1/(1+\e)}.
\label{e:u}\end{equation}
The unemployment gap $u-u^*$ can be measured from \eqref{e:u}.
\end{proposition} 

The proposition gives an explicit formula for the unemployment gap, expressed in terms of sufficient statistics.\footnote{We could obtain a formula for the unemployment gap without assumption~\ref{a:constant}, but we would need three additional statistics: the elasticities of $\e$, $\z$, and $\k$ with respect to unemployment \cp{K18}.} The formula is valid in any Beveridgean model, irrespective of the microfoundations for the Beveridge curve, of firms' production functions, of workers' utility functions, and of wage setting. Another advantage of the formula is that we do not need to keep track of all the shocks disturbing the labor market; we only need to keep track of the sufficient statistics.

\section{Application to the DMP model}\label{s:dmp}

We now apply sufficient-statistic formula \eqref{e:theta} to the most widely used Beveridgean model---the DMP model. In particular, we compare the efficient allocation given by our formula to that given by the well-known \ct{H90} condition. We base our application on the DMP model presented by \ct[chapter~1]{P00}.

\subsection{Beveridge elasticity}

\paragraph{Matching function} We assume a Cobb-Douglas matching function:
\begin{equation}
m(u,v)=\o u^{\h}v^{1-\h},
\label{e:matching}\end{equation}
where $\o>0$ is the matching efficacy, and $\h\in(0,1)$ is the matching elasticity.

\paragraph{Unemployment dynamics} The unemployment rate evolves according to the following differential equation:
\begin{equation}
\dot{u}(t) = \l [1-u(t)] - m(u(t),v(t)),
\label{e:uDot}\end{equation}
where $\l$ is the job-separation rate. The term $\l [1-u(t)]$ gives the number of workers who lose or quit their jobs and enter unemployment during a unit time. The term $m(u(t),v(t))$ gives the number of workers who find a job and leave unemployment during a unit time. The difference between inflow and outflow determines the change in the unemployment rate, $\dot{u}$.

Equation \eqref{e:uDot} can be expressed as a linear differential equation:
\begin{equation}
\dot{u}(t) + (\l+f) [u(t)- u^b] = 0,
\label{e:uDotLinear}\end{equation}
where $f = m(u,v)/u = \o \t^{1-\h}$ is the job-finding rate, and 
\begin{equation}
u^b = \frac{\l}{\l+f}
\label{e:ub}\end{equation}
is the Beveridgean unemployment rate---the critical point of the differential equation. At the Beveridgean unemployment rate, the inflow into unemployment equals the outflow from unemployment. To solution to differential equation~\eqref{e:uDotLinear} is
\begin{equation}
u(t) - u^b = [u(0) - u^b] e^{-(\l+f)t}.
\label{e:ut}\end{equation}

\paragraph{Half-life of the deviation from Beveridgean unemployment} Equation \eqref{e:ut} shows that the distance between the unemployment rate $u(t)$ and the Beveridgean unemployment rate $u^b$ decays at an exponential rate $\l+f$. In the United States, the rate of decay is really fast: between 1951 and 2019, the job-finding rate averages $f = 58.7\%$ per month, the job-separation rate averages $\l = 3.4\%$ per month, so the rate of decay averages $\l+f = 62.1 \%$ per month (appendix~\ref{a:distance}). Accordingly, the half-life of the deviation from the Beveridgean unemployment rate, $u(t)-u^b$, is $\ln(2)/0.621 = 1.1$ month. Since about $50\%$ of the deviation evaporates within one month, and about $90\%$ within one quarter, the unemployment rate is always close the Beveridgean unemployment rate---as already noted by \ct[p.~88]{EMS09}.\footnote{Appendix~\ref{a:distance} computes the Beveridgean unemployment rate in the United States and confirms that it is almost indistinguishable from the actual unemployment rate.}

\paragraph{Beveridge curve} Given such short half-life, it is accurate to assume that the inflow into unemployment equal the outflow from unemployment at all times: $\l (1-u) = m(u,v)$. Then the labor market is always on the Beveridge curve
\begin{equation}
v(u) = \bp{\frac{\l}{\o}\cdot\frac{1-u}{u^{\h}}}^{1/(1-\h)}.
\label{e:v}\end{equation}
The function $v(u)$ satisfies assumption~\ref{a:v}; in particular, it is strictly convex (appendix~\ref{a:proofs}).

\paragraph{Beveridge elasticity} From the Beveridge curve \eqref{e:v}, we obtain the Beveridge elasticity:
\begin{equation}
\e = \frac{1}{1-\h}\bp{\h+\frac{u}{1-u}}.
\label{e:epsilonDmp}\end{equation}
The Beveridge elasticity is closely related to the matching elasticity, $\h$.

Unlike what assumption~\ref{a:constant} postulates, the Beveridge elasticity depends on the unemployment rate, $u$. However, in practice, the unemployment rate is an order of magnitude smaller than the matching elasticity, so the term $u/(1-u)$ is an order of magnitude smaller than the term $\h$. Hence, the Beveridge elasticity is approximately independent from the unemployment rate.\footnote{Appendix~\ref{a:endogeneity} shows that in a DMP model calibrated to US data, the endogeneity of the Beveridge elasticity has almost no effect on the efficient unemployment rate. The efficient unemployment rate obtained with the elasticity \eqref{e:epsilonDmp} never deviates by more than $0.18$ percentage point from the baseline obtained with a constant elasticity.}

\subsection{Social value of nonwork and recruiting cost}

The welfare function is given by \eqref{e:w}. From it, we see that the social value of nonwork $\z$ and recruiting cost $\k$ correspond to parameters of the model:
\begin{equation}
\z = z \quad\text{and}\quad \k = c.
\label{e:zkdmp}\end{equation}

\subsection{Efficiency condition}

We now use \eqref{e:epsilonDmp} and \eqref{e:zkdmp} to express efficiency condition \eqref{e:theta} in terms of parameters of the DMP model:
\begin{equation*}
\t =  \frac{1-\h}{\h+u/(1-u)}\cdot \frac{1-z}{c}.
\end{equation*}
On the Beveridge curve, labor flows are balanced, so $\l (1-u) = f(\t) u $, where $f(\t) = \o (\t)^{1-\h}$ is the job-finding rate. This means that $u/(1-u) =\l/f(\t)$. Accordingly, the efficient tightness $\t^*$ is implicitly defined by
\begin{equation}
\h\t^* + \frac{\l}{q(\t^*)} = (1-\h) \frac{1-z}{c},
\label{e:thetaDmp}\end{equation}
where $q(\t) = \o \t^{-\h}$ is the vacancy-filling rate.\footnote{The left-hand side of \eqref{e:thetaDmp} is continuous and strictly increasing from $0$ to $\infty$ when $\t^*$ goes from $0$ to $\infty$. Since the right-hand side is a positive number, equation \eqref{e:thetaDmp} admits a unique solution.}

\subsection{Comparison with the Hosios condition}

In the DMP model, workers negotiate their wages with firms via Nash bargaining. When workers' bargaining power $\b$ equals the matching elasticity $\h$, the labor market is guaranteed to operate efficiently \cp{H90}.\footnote{\ct[p.~281]{H90} proves the result by assuming that the discount rate is zero and therefore that the social planner maximizes steady-state welfare. But the result continues to hold when the discount rate is positive and the social planner maximizes the present-discounted sum of flow social welfare \cp[pp.~183--185]{P00}.} We now compare the tightness $\t^*$ given by efficiency condition \eqref{e:thetaDmp} with the tightness $\t^h$ stemming from the Hosios condition. These two tightnesses may differ because they solve different planning problems: in our planning problem the Beveridge curve holds at all times, whereas in the Hosios planning problem the unemployment rate follows differential equation \eqref{e:uDotLinear}.

In the DMP model, tightness is determined by the job-creation curve
\begin{equation}
(1-\b)(1-z) - \bs{\frac{r+\l}{q(\t)} + \b\t} c = 0,
\label{e:jobCreation}\end{equation}
where $r$ is the discount rate \cp[equation~(1.24)]{P00}. This expression holds even if the labor market is temporarily away from the Beveridge curve \cp[pp.~26--32]{P00}. It is obtained by combining the wage equation, which describes the wages obtained by Nash bargaining, and the free-entry condition, which says that vacancies are created until all profit opportunities are exploited. When the Hosios condition holds, $\b=\h$, so the tightness $\t^h$ satisfies
\begin{equation}
\h\t^h + \frac{r+\l}{q(\t^h)} = (1-\h)\frac{1-z}{c}.
\label{e:thetaHosios}\end{equation}

Comparing \eqref{e:thetaHosios} with \eqref{e:thetaDmp}, we find that the tightnesses $\t^*$ and $\t^h$ are almost identical:

\begin{proposition}\label{p:hosios} In the DMP model with zero discount rate, the tightness $\t^*$ given by efficiency condition \eqref{e:theta} is the same as the tightness $\t^h$ given by the Hosios condition. In the DMP model with positive discount rate, the two tightnesses are different, but the difference is minor for realistic discount rates. To a first-order approximation, the relative deviation between the two tightnesses is
\begin{equation*}
\frac{\t^*-\t^h}{\t^*} = \frac{r}{\h (\l+f)}. 
\end{equation*}
Under the calibration in \ct[table~2]{S05}, 
\begin{equation*}
\frac{\t^*-\t^h}{\t^*} = 1.1\%. 
\end{equation*}\end{proposition}

The proof is relegated to appendix~\ref{a:proofs}, but the intuition is simple. When the discount rate is zero, equations \eqref{e:thetaDmp} and \eqref{e:thetaHosios} coincide, so they give the same tightness. When the discount rate is positive, the two tightnesses differ; but the difference just like the discount rate is small.\footnote{In appendix~\ref{a:hosios}, we simulate a DMP model to compare the unemployment rate given by the sufficient-statistic formula \eqref{e:u} with the unemployment rate implied by the Hosios condition. We find that the two unemployment rates are close: the average absolute distance is only $0.17$ percentage point.}

\begin{figure}[t!]
\subcaptionbox{Efficient unemployment rate\label{f:dmpZero}}{\includegraphics[scale=\sfig,page=12]{\pdf}}\hfill
\subcaptionbox{Positive unemployment gap\label{f:dmpPositive}}{\includegraphics[scale=\sfig,page=13]{\pdf}}\vfig
\subcaptionbox{Negative unemployment gap\label{f:dmpNegative}}{\includegraphics[scale=\sfig,page=14]{\pdf}}
\caption{Efficient unemployment rate and unemployment gap in the DMP model}
\note{The Beveridge curve is given by \eqref{e:v}. The isowelfare curve has slope $-(1-z)/c$, where $z$ is the relative productivity of unemployment workers, and $c$ is the recruiting cost. The job-creation curve is given by \eqref{e:jobCreation}. 
The point of tangency between the Beveridge curve and isowelfare curve gives the efficient allocation. The intersection of the Beveridge curve and job-creation curve gives the solution of the DMP model.}
\label{f:dmp}\end{figure}

\subsection{Efficiency in the Beveridge diagram}

We now illustrate the efficiency properties of the DMP model in a Beveridge diagram (figure~\ref{f:dmp}). 

\paragraph{Efficiency condition} We first plot the Beveridge curve of the DMP model, given by \eqref{e:v}. To find the efficient allocation, we add an isowelfare curve. From \eqref{e:zkdmp}, we see that the isowelfare curves are linear with slope $-(1-z)/c$. The efficient allocation is the point on the Beveridge curve that is tangent to an isowelfare curve.

The solution of the DMP model is given by the intersection of the Beveridge curve and job-creation curve \eqref{e:jobCreation}. Since the job-creation curve determines a tightness $\t$, without involving unemployment and vacancies, it is represented by a ray through the origin of slope~$\t$. 

When the labor market operates efficiently, the job-creation curve runs through the efficiency point on the Beveridge curve (figure~\ref{f:dmpZero}). When the discount rate is zero, our efficiency condition coincides with the Hosios condition, so workers' bargaining power in the job-creation curve is $\b = \h$. When the discount rate is positive, the two conditions differ but the difference is minuscule, so $\b \approx \h$.

\paragraph{Deviations from efficiency} The labor market may not operate efficiently, however. If workers' bargaining power is too high, the job-creation curve is too low: unemployment is too high, vacancies are too low, and the unemployment gap is positive (figure~\ref{f:dmpPositive}). Conversely, if workers' bargaining power is too low, the job-creation curve is too high: unemployment is too low, vacancies are too high, and the unemployment gap is negative (figure~\ref{f:dmpNegative}).

\section{Unemployment gap in the United States, 1951--2019}\label{s:usa}

Next we use sufficient-statistic formula \eqref{e:u} to measure the unemployment gap in the United States between 1951 and 2019. The first step is to estimate the sufficient statistics: Beveridge elasticity, social value of nonwork, and recruiting cost.

\subsection{Beveridge elasticity} 

We estimate the Beveridge elasticity in the United States by regressing log vacancy rate (from figure~\ref{f:v}) on log unemployment rate (from figure~\ref{f:u}). The data are quarterly from 1951Q1 to 2019Q4, so the sample contains $276$ observations. Since the Beveridge curve shifts at multiple points in time, we use the algorithm proposed by \ct{BP98,BP03} to estimate linear models with multiple structural breaks.

\paragraph{Statistical model}  We consider a statistical model with $m$ breaks and $m+1$ regimes:
\begin{equation*}
\ln(v(t)) = \ln(\a_j) - \e_j \cdot \ln(u(t)) + z(t)  \qquad t = T_{j-1}+1, \ldots,T_j
\end{equation*} 
for $j = 1, \ldots, m+1$. The observed dependent variable is the log vacancy rate, $\ln(v(t))$. The observed independent variable is the log unemployment rate, $\ln(u(t))$. The error term is denoted $z(t)$. The $m$ break dates are denoted $T_1,\ldots,T_m$; moreover, $T_0=0$ and $T_{m+1}=276$. The parameter $\a_j$ determines the intercept of the linear model in regime $j$. Finally, the parameter $\e_j$ is the Beveridge elasticity in regime $j$. 

We jointly estimate the parameters $\a_1,\ldots,\a_{m+1}$ and $\e_1,\ldots,\e_{m+1}$ and the break dates $T_1,\ldots,T_m$ with the Bai-Perron algorithm. The algorithm first determines the number of structural breaks, $m$. It then estimates the parameters by least-squares and the break dates by minimizing the sum of squared residuals. It finally computes confidence intervals for the parameters and break dates.

\begin{figure}[p]
\subcaptionbox{First branch: 1951Q1--1961Q1\label{f:regime1}}{\includegraphics[scale=\sfig,page=15]{\pdf}}\hfill
\subcaptionbox{Second branch: 1961Q2--1971Q4\label{f:regime2}}{\includegraphics[scale=\sfig,page=16]{\pdf}}\vfig
\subcaptionbox{Third branch: 1972Q1--1989Q1\label{f:regime3}}{\includegraphics[scale=\sfig,page=17]{\pdf}}\hfill
\subcaptionbox{Fourth branch: 1989Q2--1999Q2\label{f:regime4}}{\includegraphics[scale=\sfig,page=18]{\pdf}}\vfig
\subcaptionbox{Fifth branch: 1999Q3--2009Q3\label{f:regime5}}{\includegraphics[scale=\sfig,page=19]{\pdf}}\hfill
\subcaptionbox{Sixth branch: 2009Q4--2019Q4\label{f:regime6}}{\includegraphics[scale=\sfig,page=20]{\pdf}}
\caption{Beveridge-curve branches in the United States, 1951--2019}
\note{The Beveridge curve comes from figure~\ref{f:beveridge}. The number of structural breaks and their dates are estimated with the algorithm of \ct{BP98,BP03}. The 5 break dates delineate 6 Beveridge-curve branches. The wide, transparent lines depict the $95\%$ confidence intervals for the break dates.}
\label{f:regimes}\end{figure}

\paragraph{Algorithm setup} We begin by setting up the Bai-Perron algorithm. First, we allow for autocorrelation in the errors, and for different variances of the errors across regimes.\footnote{Temporary deviations from the balanced-flow assumption in the data appear in the errors. This is why we allow the errors to be autocorrelated and heteroskedastic. \ct{AC20} impose more structure on labor-market flows, which allows them to quantify the deviations from the balanced-flow assumption and to estimate the Beveridge elasticity more finely.} To obtain standard errors robust to autocorrelation and heteroskedasticity, the algorithm uses a quadratic kernel with automatic bandwidth selection based on an AR(1) approximation, as proposed by \ct{A91}. Second, we allow for different distributions of the independent and dependent variables across regimes. Third, we set the trimming parameter to $0.15$, as suggested by \ct[p.~15]{BP03}; hence each regime has at least $0.15 \times 276 = 41$ observations. Fourth, we set the maximum number of breaks to $5$, as required by \ct[p.~14]{BP03}.

\paragraph{Number of breaks} Next, we determine the number of structural breaks. We first examine whether a structural break exists. We run supF tests of no structural break versus $m$ breaks, for $m=1,\ldots,5$. The tests reject the null hypothesis of no break at the $1\%$ significance level. We also run double-maximum tests of no structural break versus an unknown number of breaks. Again, the tests reject the null hypothesis of no break at the $1\%$ significance level. Thus, at least one break is present. To estimate the number of breaks, we consider two information criteria: the Bayesian Information Criterion proposed by \ct{Y88}, and the modified Schwarz criterion proposed by \ct{LWZ97}. Both criteria select $5$ breaks. 

\paragraph{Break dates} Next we estimate the $5$ break dates. We find that the breaks occur in 1961Q1, 1971Q4, 1989Q1, 1999Q2, and 2009Q3. The break dates are precisely estimated as all their $95\%$ confidence intervals cover less than $2.5$ years. The $6$ branches of the Beveridge curve delineated by the break dates, together with the $95\%$ confidence intervals for the dates, are represented in figure~\ref{f:regimes}.

\paragraph{Beveridge elasticity} Finally, we estimate the Beveridge elasticity in the 6 different regimes. The elasticity estimates remain between $0.84$ and $1.02$, averaging $0.91$ over 1951--2019 (figure~\ref{f:epsilon}). The elasticity estimates are fairly precise: the standard errors remain between $0.06$ and $0.15$. The fit of the model is $R^2 = 0.91$. Such good fit confirms that unemployment and vacancy rates travel in the vicinity of an isoelastic curve that occasionally shifts.

\begin{figure}[t]
\includegraphics[scale=\sfig,page=21]{\pdf}
\caption{Beveridge elasticity in the United States, 1951--2019}
\note{The Beveridge elasticity (thick purple line) is estimated by applying the \ct{BP98,BP03} algorithm to the log vacancy and unemployment rates from figure~\ref{f:beveridge}. The $95\%$ confidence interval for the elasticity (purple area) is computed using standard errors corrected for autocorrelation in the error terms, as well as heterogeneity in the data and in the error terms across regimes. The shaded areas are NBER-dated recessions.}
\label{f:epsilon}\end{figure}

\paragraph{Comparison with estimates of the matching elasticity} In the DMP model, the Beveridge elasticity is directly related to the matching elasticity. Using \eqref{e:epsilonDmp}, we find that our estimates of the Beveridge elasticity translate into estimates of the matching elasticity between $0.39$ and $0.49$, with an average of $0.44$ over 1951--2019 (figure~\ref{f:eta}). These estimates of the matching elasticity fall squarely in the range of estimates obtained with aggregate US data, which spans $0.30$--$0.76$.\footnote{\ct[table~1]{BD89} obtain estimates of the matching elasticity between $0.32$ and $0.60$. \ct[table~1]{BlF97} obtain estimates between $0.54$ and $0.76$. \ct[p.~32]{S05} obtains an estimate of $0.72$. \ct[p.~638]{RS10} obtain an estimate of $0.58$. Last, \ct[p.~444]{BJP11} obtain an estimate of $0.30$.}

\subsection{Social value of nonwork}

We measure the social value of nonwork in the United States from revealed-preference estimates.

\paragraph{Revealed-preference estimates} Using administrative data from the US military, \ct{BM15} study how servicemembers choose between reenlisting and leaving the military. The choices allow them to estimate the utility loss caused by unemployment during the transition to civilian life. They find that home production, recreation, and public benefits during unemployment offset between $13\%$ and $35\%$ of the earnings loss caused by unemployment.

Using a large field experiment in the United States, \ct{MP19} study how unemployed job applicants choose between randomized wage-hour bundles. The choices imply that home production and recreation during unemployment are worth $58\%$ of predicted earnings. 

\paragraph{Translating revealed-preference estimates into social values of nonwork} Next, we translate these estimates into social values of nonwork. We ignore the fact that employed and unemployed workers value consumption differently, which allows us to measure workers' contribution to welfare directly from their productivity, at home or at work.\footnote{\ct{LS20} provide revealed-preference evidence on the difference between the marginal values of consumption for unemployed and employed workers. This evidence could be combined with the methodology of \ct{LMS15} to measure the social value of nonwork when unemployed workers are imperfectly insured.} For unemployed workers, the contribution to social welfare should not include unemployment benefits, which are just transfers.

We begin by expressing the estimates relative to the marginal product of labor rather than to earnings. The marginal product of labor is higher than earnings for several reasons. First, firms usually pay less than the marginal product of labor. In a matching model, the marginal product of labor is about $3\%$ higher than the wage \cp[equation~(1)]{LMS10}. In a monopsony model, the marginal product of labor may be $25\%$ higher than the wage \cp[p.~121]{MP19}. Second, in the United States, workers earn less than the wage paid by firms because of the $7.7\%$ employer-side payroll tax. Third, \name{MP19} discount predicted earnings by $6\%$ to capture the wage penalty incurred by workers who recently lost their jobs; we undo the discounting because the penalty does not apply to the marginal product of labor \cp{DV11}. To conclude, to obtain a marginal product of labor, \name{BM15}'s earnings must be adjusted by a factor between $1.03\times 1.077 = 1.11$ and $1.25 \times 1.077 = 1.35$, and \name{MP19}'s earnings must be adjusted by a factor between $1.03\times 1.077 \times 1.06 = 1.18$ and $1.25 \times 1.077 \times 1.06 = 1.43$. Accordingly, to obtain an estimate relative to the marginal product of labor, \name{BM15}'s numbers must be adjusted by a factor between $1/1.35 = 0.74$ and $1/1.11 = 0.90$, and \name{MP19}'s number must be adjusted by a factor between $1/1.43 = 0.70$ and $1/1.18 = 0.85$. 

Additionally, we subtract the value of public benefits from \name{BM15}'s estimates. All servicemembers are eligible to unemployment insurance (UI). \ct[pp.~1585--1586]{CK16} find that UI benefits amount to $21.5\%$ of the marginal product of labor. But this quantity has to be reduced for several reasons: the UI takeup rate is only $65\%$; UI benefits and consumption are taxed, imposing a factor of $0.83$; the disutility from filing for benefits imposes a factor of $0.47$; and UI benefits expire, imposing another factor of $0.83$. In sum, the average value of UI benefits to servicemembers is $21.5 \times 0.65 \times 0.83 \times 0.47 \times 0.83 = 5\%$ of the marginal product of labor. Servicemembers are also eligible to other public benefits, which \name{CK16} quantify at $2\%$ of the marginal product of labor. Hence, to account for benefits, we subtract $5\%+2\% =7\%$ of the marginal product of labor from \name{BM15}'s estimates.

Overall, we find that \name{BM15}'s estimates imply a social value of nonwork between $(0.13 \times 0.74)-0.07 = 0.03$ and $(0.35 \times 0.90)-0.07 = 0.25$, and that \name{MP19}'s estimates imply a social value of nonwork between $0.58 \times 0.70 = 0.41$ and $0.58 \times 0.85 = 0.49$. The range of plausible social values of nonwork therefore is $0.03$--$0.49$. We set the statistic to its midrange value: $\z=0.26$.

\paragraph{Fluctuations in the social value of nonwork} In some models, the productivities of unemployed and employed workers do not move in tandem over the business cycle, which generates fluctuations in the social value of nonwork. However, \ct[pp.~1599--1604]{CK16} find no evidence of such fluctuations in US data. Instead, they establish that the value of home production and recreation during unemployment moves proportionally to labor productivity---which implies that the social value of nonwork is acyclical. Accordingly, we keep the social value of nonwork constant over the business cycle.\footnote{In any case, appendix~\ref{a:fluctuations} shows that the efficient unemployment rate is virtually unchanged if the social value of nonwork fluctuates in response to variations in labor productivity.} The social value of nonwork could also exhibit medium-run fluctuations; we omit them by lack of evidence.

\paragraph{Other contributors to the social value of nonwork} Here we measure the social value of nonwork by revealed preferences. This approach captures the value of nonwork that transpires from people's choices, but it might miss externalities imposed by nonwork.

For instance, higher unemployment might lead to more crime. If the externality was strong, the social value of nonwork would be less than that given by the revealed-preference approach. However, unemployment appears to stimulate crime only weakly \cp{Fr99}. Hence, as a first pass, we do not deduct the cost of crime from our estimate of the social value of nonwork.

Higher unemployment might also impede upward mobility and raises inequality \cp{O73,ADW19}. When the social planner dislikes inequality, this externality reduces the social value of nonwork. As a first pass, we ignore the externality; but it could be included in future estimates of the social value of nonwork.

\subsection{Recruiting cost}

Following \ct{V10}, we measure the recruiting cost in the United States from the National Employer Survey conducted by the \ct{NES} in 1997. The survey asked thousands of establishments across industries about their recruiting practices \cp{Cap01}. In the public-use files of the survey, $2007$ establishments report the fraction of labor costs devoted to recruiting; the mean response is $3.2\%$. If all workers are paid the same, the establishments allocate $3.2\%$ of their labor to recruiting, so $\k v = 3.2\% \times (1-u)$. In 1997, the vacancy rate is $3.3\%$ and the unemployment rate is $4.9\%$ (figure~\ref{f:beveridge}). Hence, the recruiting cost in 1997 is $\k =  3.2\% \times (1-4.9\%) / 3.3\% = 0.92$.

We do not know how the US recruiting cost varies over time because there is no other comprehensive measure of it. However, in matching models, the recruiting cost is usually assumed to be constant over time. Lacking evidence, we follow this tradition and assume that the recruiting cost remains at its 1997 value from 1951 to 2019. 

This lack of evidence is not ideal to measure past unemployment gaps, but it could be remedied in the future by adding a question to the Job Opening and Labor Turnover Survey. The recruiting cost could be measured every month by asking firms to report how many man-hours they devote to recruiting in addition to their number of vacancies.

\subsection{Unemployment gap}

We now use our estimates of the Beveridge elasticity, social value of nonwork, and recruiting cost to measure the unemployment gap in the United States from 1951 to 2019.

\begin{figure}[t!]
\subcaptionbox{Efficient labor-market tightness\label{f:thetaEfficient}}{\includegraphics[scale=\sfig,page=22]{\pdf}}\hfill
\subcaptionbox{Efficient unemployment rate\label{f:uEfficient}}{\includegraphics[scale=\sfig,page=23]{\pdf}}\vfig
\subcaptionbox{Unemployment gap\label{f:uGap}}{\includegraphics[scale=\sfig,page=24]{\pdf}}\hfill
\subcaptionbox{Alternative unemployment rates\label{f:gaps}}{\includegraphics[scale=\sfig,page=25]{\pdf}}
\caption{Unemployment gap in the United States, 1951--2019}
\note{A: The efficient tightness is computed using \eqref{e:theta} with $\e$ from figure~\ref{f:epsilon}, $\z = 0.26$, and $\k=0.92$. The actual tightness is the vacancy rate from figure~\ref{f:v} divided by the unemployment rate from figure~\ref{f:u}; it is displayed as a benchmark. B: The efficient unemployment rate is computed using \eqref{e:u} with $\e$ from figure~\ref{f:epsilon}, $\z = 0.26$, $\k=0.92$, and the unemployment and vacancy rates from figure~\ref{f:beveridge}. The actual unemployment rate comes from figure~\ref{f:u}; it is displayed as a benchmark. C: The unemployment gap is the difference between the actual unemployment rate and the efficient unemployment rate from panel~B. D: The efficient unemployment rate comes from panel~B; NAIRU and trend unemployment rate come from \ct[figure~8B]{CEG19}; the natural unemployment rate is constructed by the \ct{NROU}. The shaded areas are NBER-dated recessions.}
\label{f:gap}\end{figure}

\paragraph{Efficient labor-market tightness} We begin by computing the efficient tightness from formula \eqref{e:theta}. The efficient tightness fluctuates between $0.79$ and $0.96$, mirroring the fluctuations of the Beveridge elasticity, and it averages $0.89$ (figure~\ref{f:thetaEfficient}).

\paragraph{Efficient unemployment rate} Next we compute the efficient unemployment rate from formula \eqref{e:u}. The efficient unemployment rate averages $4.3\%$, and it always remains between $3.0\%$ and $5.4\%$ (figure~\ref{f:uEfficient}). It hovered around $3.5\%$ in the 1950s and around $4.5\%$ in the 1960s, and it climbed to $5.4\%$ at the end of the 1970s. This steady increase was caused by a steady outward shift of the Beveridge curve (figures \ref{f:regime1}--\ref{f:regime3}). The efficient unemployment rate then declined to $4.6\%$ at the end of the 1980s. The decline was caused by an inward shift of the Beveridge curve (figures \ref{f:regime3}--\ref{f:regime4}). Last, the efficient unemployment rate remained stable through the 1990s, 2000s, and 2010s, hovering between $3.8\%$ and $4.6\%$. The efficient unemployment rate did not increase after the Great Recession, despite an outward shift of the Beveridge curve (figures \ref{f:regime5}--\ref{f:regime6}). This is because the Beveridge curve also became flatter after 2009: the Beveridge elasticity fell from $1.0$ to $0.84$ (figure~\ref{f:epsilon}). The flattening offset the outward shift, leaving the efficient unemployment rate almost unchanged.

\paragraph{Unemployment gap} We finally compute the unemployment gap by subtracting the efficient unemployment rate from the actual unemployment rate (figure~\ref{f:uGap}). First, the unemployment gap is almost never zero, so the US labor market is almost never efficient. Second, the unemployment gap is almost always positive, so the US labor market is almost always inefficiently slack. The unemployment gap averages $1.4$ percentage points over 1951--2019. And it was only negative during four episodes: 1951--1953, during the Korean war; 1965--1970, at the peak of the Vietnam war; 1999--2000, during the dot-com bubble; and 2018--2019. Third, the unemployment gap is sharply countercyclical, so inefficiencies are exacerbated in slumps. The unemployment gap is close to zero at business-cycle peaks: $0.4$ percentage point in 1979, $-0.3$ percentage point in 2000, or $0.3$ percentage point in 2007. But it is highly positive at business-cycle troughs: $6.1$ percentage  points in 1982, $3.2$ percentage points in 1992, or $6.2$ percentage points in 2009. Unsurprisingly, the largest unemployment gaps occurred after the Volcker Recession and after the Great Recession.

\subsection{Relaxing assumption~\ref{a:constant}}

Although assumption~\ref{a:constant} is required to compute the unemployment gap, we can determine whether unemployment is inefficiently high or low without it. Indeed, unemployment is inefficiently high whenever tightness is inefficiently low, which happens whenever $\t < (1-\z)/(\k\e)$ (proposition~\ref{p:theta}). Since \eqref{e:u} can be rewritten
\begin{equation*}
\frac{u^*}{u} = \bs{\frac{\t}{(1-\z)/(\k \e)} }^{1/(1+\e)},
\end{equation*}
we infer that $\t < (1-\z)/(\k\e)$ whenever $u^*<u$. That is, unemployment is inefficiently high whenever the unemployment gap in figure~\ref{f:uGap} is positive; conversely, unemployment is inefficiently low whenever the unemployment gap in figure~\ref{f:uGap} is negative. In other words, without assumption~\ref{a:constant}, the size of the unemployment gap in figure~\ref{f:uGap} may be inaccurate, but its sign is valid.

\subsection{Comparisons with other unemployment gaps} 

To provide some context, we compare our efficient unemployment rate to other unemployment rates that are commonly used to construct unemployment gaps: unemployment-rate trend, NAIRU, and natural rate of unemployment (figure~\ref{f:gaps}). The unemployment-rate trend and NAIRU are constructed by \ct[figure~8B]{CEG19} using state-of-the-art techniques. The natural rate of unemployment is constructed by the \ct{NROU}.

Although these unemployment rates feature prominently in policy discussions, they are not designed to measure efficiency. In most models unemployment is not efficient on average, so the unemployment-rate trend cannot mark efficiency \cp{Hall05}. The NAIRU is obtained by estimating a Phillips curve, so it is not meant to indicate labor-market efficiency \cp{R97}. The natural rate of unemployment blends trend and NAIRU considerations \cp{Sh18}; it does not indicate efficiency either.

Nevertheless, the four unemployment series share similarities. First, the four series are slow-moving. As the actual unemployment rate is sharply countercyclical, the four series produce countercyclical unemployment gaps. Another similarity is that the four unemployment series were higher in the 1970s and 1980s, and lower after that. 

The main difference is that our series is lower than the three others. On average the efficient unemployment rate is $1.5$ percentage points below the unemployment-rate trend, $1.2$ percentage points below the NAIRU, and $1.6$ percentage points below the natural rate of unemployment. As a result, the unemployment gap constructed with the efficient unemployment rate is higher than that constructed with the other series. However, the four series converge in the 2010s, and as of 2018, they are close: between $3.8\%$ and $4.5\%$.

\subsection{Alternative calibrations of the sufficient statistics}

\begin{figure}[t!]
\subcaptionbox{Beveridge elasticity: $\e$ in $95\%$ confidence interval\label{f:rangeEpsilon}}{\includegraphics[scale=\sfig,page=26]{\pdf}}\hfill
\subcaptionbox{Social value of nonwork: $0.03 < \z < 0.49$\label{f:rangeZeta}}{\includegraphics[scale=\sfig,page=27]{\pdf}}\vfig
\subcaptionbox{Recruiting cost: $0.61 <\k < 1.23$\label{f:rangeKappa}}{\includegraphics[scale=\sfig,page=28]{\pdf}}
\caption{US efficient unemployment rate for a range of sufficient statistics}
\note{The figure reproduces figure~\ref{f:uEfficient}, and adds the ranges of efficient unemployment rates obtained when the sufficient statistics span all plausible values (pink areas). A: The $95\%$ confidence interval of $\e$ comes from figure~\ref{f:epsilon}. The bottom pink line corresponds to the bottom-end value of $\e$, and the top pink line to the top-end value of $\e$. B: The bottom pink line corresponds to $\z=0.03$, and the top pink line to $\z=0.49$. C: The bottom pink line corresponds to $\k=0.61$, and the top pink line to $\k=1.23$.}
\label{f:range}\end{figure}

To assess the robustness of our findings, we explore the sensitivity of the efficient unemployment rate to alternative calibrations of the sufficient statistics. We also compute the inverse-optimum values of the sufficient statistics---which ensure that the labor market is always efficient \cp{H20}. The distance between the inverse-optimum and calibrated statistics is another measure of labor-market inefficiency.

\paragraph{Beveridge elasticity} We construct the efficient unemployment rate when the Beveridge elasticity takes any value in its 95\% confidence interval (figure~\ref{f:rangeEpsilon}). When the elasticity is at the bottom end of the confidence interval, the efficient unemployment rate follows the same pattern as under the baseline calibration but is on average $0.6$ percentage point lower (bottom pink line). When the elasticity is at the top end of the confidence interval, the efficient unemployment rate follows the same pattern as under the baseline calibration but is on average $0.5$ percentage point higher (top pink line). For any elasticity inside the confidence interval, the efficient unemployment rate is somewhere between these two extremes (pink area). The width of the pink area shows that for any elasticity inside the confidence interval, the efficient unemployment rate never deviates by more than $1.2$ percentage points from its baseline.

Next we compute the inverse-optimum Beveridge elasticity from proposition~\ref{p:theta}. Given the other sufficient statistics, actual tightness is efficient if the Beveridge elasticity is
\begin{equation}
\e^* = \frac{1-\z}{\k\t}.
\label{e:epsilon}\end{equation}
The inverse-optimum Beveridge elasticity $\e^*$ is strongly countercyclical: it varies between $0.5$ in booms and $5.0$ during the Great Recession, with an average value of $1.6$ (figure~\ref{f:inverseEpsilon}). It is generally above the $95\%$ confidence interval of the estimated Beveridge elasticity, and very far above it in slumps---confirming that the labor market is inefficiently slack in slumps.

\begin{figure}[t!]
\subcaptionbox{Inverse-optimum Beveridge elasticity\label{f:inverseEpsilon}}{\includegraphics[scale=\sfig,page=30]{\pdf}}\hfill
\subcaptionbox{Inverse-optimum social value of nonwork\label{f:inverseZeta}}{\includegraphics[scale=\sfig,page=31]{\pdf}}\vfig
\subcaptionbox{Inverse-optimum recruiting cost\label{f:inverseKappa}}{\includegraphics[scale=\sfig,page=32]{\pdf}}
\caption{Inverse-optimum sufficient statistics in the United States, 1951--2019}
\note{A: The inverse-optimum Beveridge elasticity is obtained from \eqref{e:epsilon}. The calibrated Beveridge elasticity comes from figure~\ref{f:epsilon}; it is displayed as a benchmark. B: The inverse-optimum social value of nonwork is obtained from \eqref{e:zeta}. The calibrated social value of nonwork is $\z=0.26$, with a plausible range $0.03$--$0.49$; it is displayed as a benchmark. C: The inverse-optimum recruiting cost is obtained from \eqref{e:kappa}. The calibrated recruiting cost is $\k= 0.92$, with a plausible range $0.61$--$1.23$; it is displayed as a benchmark. The shaded areas are NBER-dated recessions.}
\label{f:inverse}\end{figure}

\paragraph{Social value of nonwork} Next we construct the efficient unemployment rate when the social value of nonwork spans the range of values given by \ct{BM15} and \ct{MP19}: $0.03 < \z < 0.49$ (figure~\ref{f:rangeZeta}). When $\z=0.03$, the efficient unemployment rate follows the same pattern as under the baseline calibration but is on average $0.6$ percentage point lower (bottom pink line). When $\z=0.49$, the efficient unemployment rate follows again the same pattern as under the baseline calibration but is on average $0.9$ percentage point higher (top pink line). In fact, for any $\z$ between $0.03$ and $0.49$, the efficient unemployment rate never deviates by more than $1.2$ percentage points from its baseline. This is reassuring as the range of plausible social values of nonwork is quite wide.

Next we compute the inverse-optimum social value of nonwork from proposition~\ref{p:theta}. Actual tightness is efficient if the social value of nonwork is
\begin{equation}
\z^* = 1-\k\e\t.
\label{e:zeta}\end{equation}
The inverse-optimum social value of nonwork $\z^*$ is immensely countercyclical: as low as $-0.32$ in booms and as high as $0.88$ during the Great Recession, with an average value of $0.48$ (figure~\ref{f:inverseZeta}). Under the inverse-optimum social value of nonwork, recessions are mere vacations.

\paragraph{Recruiting cost} We do not have enough evidence to construct a plausible range of recruiting costs. Instead we construct an artificial range, which we use to assess the sensitivity of the efficient unemployment rate to the recruiting cost (figure~\ref{f:rangeKappa}). We consider recruiting costs between two thirds and four thirds of our estimate, so between $\k = 2/3\times 0.92 = 0.61$ and $\k = 4/3\times 0.92 = 1.23$. When $\k=0.61$, the efficient unemployment rate follows the same pattern as under the baseline calibration but is on average $0.8$ percentage point lower (bottom pink line). When $\k=1.23$, the efficient unemployment rate follows the same pattern as under the baseline calibration but is on average $0.7$ percentage point higher (top pink line). For any $\k$ between $0.61$ and $1.23$, the efficient unemployment rate remains within $1.1$ percentage points of its baseline.

Finally, we compute the inverse-optimum recruiting cost from proposition~\ref{p:theta}. Actual tightness is efficient if the recruiting cost is
\begin{equation}
\k^* = \frac{1-\z}{\e\t}.
\label{e:kappa}\end{equation}
The inverse-optimum recruiting cost $\k^*$ is strongly countercyclical, varying between $0.5$ in booms and $5.5$ during the Great Recession, with an average value of $1.7$ (figure~\ref{f:inverseKappa}). Hence, for unemployment fluctuations to be efficient, recruiting must require ten times more labor in slumps than in booms, which seems implausible. 

\paragraph{Conclusion} There remains clear uncertainty about the values of the sufficient statistics. Yet, for all plausible values, the unemployment gap remains within $1.2$ percentage point of its baseline. Furthermore, the inverse-optimum values of the statistics are generally far away from the calibrated values. Thus our findings---that the US labor market is generally inefficient and is inefficiently slack in slumps---seem robust.

\paragraph{Aside on some macro-labor calibrations of the social value of nonwork} We find that the social value of nonwork is much below $1$: $\z= 0.26$. In contrast, some macro-labor papers argue that it is very close to $1$. A well-known calibration, due to \ct{HagM08}, is $\z= 0.96$. Such calibration pushes the efficient unemployment rate above $14.9\%$, and sometimes as high as $26.3\%$, with an average value of $20.2\%$ (figure~\ref{f:hm}). In that case, the US unemployment rate is always inefficiently low, even at the peak of the Great Recession. This result is implausible, which implies that \name{HagM08}'s calibration overstates the social value of nonwork.

\begin{figure}[t]
\includegraphics[scale=\sfig,page=29]{\pdf}
\caption{US unemployment gap when the social value of nonwork is $\z=0.96$}
\note{The figure reproduces figure~\ref{f:uEfficient} but uses $\z=0.96$---as in \ct{HagM08}---instead of $\z = 0.26$.}
\label{f:hm}\end{figure}

\section{Summary and Implications}

To conclude, we summarize the results of the paper and discuss their implications.

\subsection{Summary}

This paper develops a new measure of the unemployment gap---the difference between the actual unemployment rate and the socially efficient unemployment rate. We consider a framework with only one structural element: a Beveridge curve relating unemployment and vacancies. This Beveridgean framework nests many modern labor-market models, including the DMP model. We find that the unemployment gap can be measured from three sufficient statistics: elasticity of the Beveridge curve, social cost of unemployment, and cost of recruiting. Applying the formula to the United States, 1951--2019, we find that the unemployment gap is countercyclical: close to zero in booms but highly positive in slumps. We infer that the US unemployment rate is generally inefficiently high, and that such inefficiency worsens in slumps.

\subsection{Implications for labor-market models}

Our Beveridgean framework is quite general: it allows for a broad range of assumptions. But the finding that US unemployment gap is sharply countercyclical is not consistent with all of them. We now discuss which assumptions can generate such countercyclical unemployment gap.

\paragraph{DMP model with Hosios condition} In the DMP model, workers' bargaining power is customarily set at the level given by the Hosios condition \cp{MP94,S05,CR08}. Such calibration is convenient because the bargaining power is difficult to estimate empirically \cp[p.~229]{P00}; but it implies that unemployment is efficient at all times. Given that the US unemployment rate is almost never efficient, the DMP model with Hosios condition might not accurately describe the labor market. This inaccuracy is particularly problematic when the model is used to design policies.

\paragraph{Models with competitive search} The DMP model assumes that search is random. An alternative assumption is that search is directed, so jobseekers target submarkets offering advantageous employment conditions. Most directed-search models then apply the competitive-search equilibrium developed by \ct{M97}. This equilibrium concept is tractable, but because it implies that the labor market is efficient, it may not be realistic.

\paragraph{DMP model with bargaining-power shocks} By contrast, a DMP model with shocks to workers' bargaining power easily generates the patterns observed during US business cycles (\inp[table~6]{S05}; \inp{JK15}). Under such shocks, the job-creation curve rotates up and down, while the Beveridge and isowelfare curves are fixed (figure~\ref{f:dmp}). Unemployment travels up and down the Beveridge curve, while the efficient unemployment rate is constant, which generates a countercyclical unemployment gap.

\paragraph{Variant of the DMP model with fixed wages} The patterns observed during US business cycles can also be generated by a variant of the DMP model with fixed wages instead of bargained wages \cp{S04,H05}. In such variant, the job-creation curve is obtained by inserting a fixed wage $w>0$ into the free-entry condition of the DMP model \cp[equation~(1.22)]{P00}:
\begin{equation}
1 - \frac{w}{p} - \frac{(r+\l)c}{q(\t)} = 0.
\label{e:jobCreationHall}\end{equation}
When labor productivity is low, the unit labor cost $w/p$ is high, so the tightness given by \eqref{e:jobCreationHall} is low (figure~\ref{f:dmpPositive}). Conversely, when labor productivity is high, the unit labor cost $w/p$ is low, so the tightness given by \eqref{e:jobCreationHall} is high (figure~\ref{f:dmpNegative}). At the same time, the Beveridge and isowelfare curves are unaffected by labor productivity. Thus, in response to labor-productivity shocks, unemployment travels up and down the Beveridge curve while the efficient unemployment rate does not change, which produces a countercyclical unemployment gap. The same results hold if wages are not fixed but merely rigid \cp{M09}.

\subsection{Implications for policy}

The countercyclicality of the US unemployment gap has numerous policy implications.

\paragraph{Distance from full employment} In the United States, the 1978 Humphrey-Hawkins Full Employment Act mandates the government to stabilize the economy at full employment. Because achieving zero unemployment is physically impossible, reaching full employment should not be interpreted as bringing unemployment to zero. Rather, it should be interpreted as reaching a socially efficient amount of unemployment. Viewed in this light, US policymakers are mandated to close the unemployment gap. They could use our unemployment-gap measure---which can be calculated in real time---to monitor how far from full employment the economy is. 

\paragraph{Optimal monetary policy} The mandate to eliminate the unemployment gap is optimal if stabilization policies have no side effects. Monetary policy is such a policy if it has no effect on inflation, or if the divine coincidence holds---such that inflation reaches its target when the unemployment gap vanishes \cp{BG07}. For instance, in a matching model with unemployment and fixed inflation, \ct{MS19} find that it is optimal to eliminate the unemployment gap by setting the nominal interest rate appropriately. As the unemployment gap is countercyclical, and a reduction in interest rate lowers unemployment \cp{BB92,C12}, it is optimal to lower the interest rate in slumps, when the unemployment gap is high, and to raise it in booms, when the unemployment gap turns negative.

When monetary policy affects inflation and the divine coincidence fails, monetary policy faces a tradeoff between closing the unemployment gap and bringing inflation to its target. It is no longer optimal to eliminate the unemployment gap, but the unemployment gap remains useful to design policy. In a New Keynesian model with unemployment and no divine coincidence, \ct[p.~23]{BG08} find that a monetary-policy rule responding to the unemployment gap and inflation achieves almost the same welfare as the optimal policy. According to that almost-optimal rule, the nominal interest rate should fall when the unemployment gap increases.

\paragraph{Optimal fiscal policy} If the government has only access to stabilization policies with side effects---policies that affect social welfare not only through unemployment but also through other channels---it is not optimal to eliminate the unemployment gap. Nevertheless, the unemployment gap remains a key determinant of optimal policy. This is because the unemployment gap measures the policy's impact on welfare through unemployment, keeping constant the policy's impact on welfare through other channels. 

Government spending is such a policy. It can reduce the unemployment gap; but in doing so, it shifts household consumption form private goods to public goods \cp{MW11}. In a model with unemployment and government spending, \ct{MS15} find that optimal government spending deviates from the \ct{S54} rule to reduce---but not eliminate---the unemployment gap. Since the unemployment gap is countercyclical, and an increase in government spending reduces unemployment \cp{Ra13}, optimal government spending is countercyclical.

\paragraph{Optimal unemployment insurance} Even policies designed to alleviate the hardship from unemployment without reducing unemployment should be adjusted when the labor market operates inefficiently. Unemployment insurance is one such policy. \ct{LMS10} show that when the labor market is inefficient, optimal unemployment insurance deviates from the \ct{B78}-\ct{C06} rule to reduce the tightness gap. Since the tightness gap is procyclical (figure~\ref{f:thetaEfficient}), and an increase in unemployment insurance raises tightness \cp[section~3]{LMS15}, optimal unemployment insurance is countercyclical.

\paragraph{Policies for disaggregated labor markets} We compute the unemployment gap for the aggregate US labor market, but the method could also be applied to local labor markets. Local unemployment gaps could be used to design local policies: to target government spending to the areas that need it most; or to tailor unemployment insurance to local labor-market conditions.

The method could also be applied to other disaggregated labor markets, such as labor markets for specific education levels. Education-specific unemployment gaps would make it possible to customize policies by education group. Hiring subsidies and firing taxes effectively modulate labor demand \cp[chapter~9]{P00}; thus, they could be tailored to eliminate education-specific unemployment gaps.

\bibliography{\bib}

\newpage
\appendix

\section{Proofs}\label{a:proofs}

This appendix provides proofs that are omitted in the main text.

\subsection{Proof that the DMP model's Beveridge curve is strictly convex}

In the DMP model, the Beveridge curve $u \mapsto v(u)$ is given by \eqref{e:v}:
\begin{equation*}
v(u) = \bp{\frac{\l}{\o}\cdot\frac{1-u}{u^{\h}}}^{1/(1-\h)}.
\end{equation*}
Since $(1-u)/u^{\h} = u^{-\h}-u^{1-\h}$, the derivative of the Beveridge curve is
\begin{equation*}
v'(u)= \frac{v(u)^{\h}}{1-\h} \cdot \frac{\l}{\o} \cdot \bs{-\h u^{-\h-1}-(1-\h) u^{-\h}}.
\end{equation*}
Reshuffling terms, we obtain
\begin{equation}
v'(u) = -\frac{\l}{\o} \cdot \bs{\frac{v(u)}{u}}^{\h}\cdot \bp{1+\frac{\h}{1-\h}\cdot\frac{1}{u}}.
\label{e:vprime}\end{equation}

From \eqref{e:vprime} we verify that the Beveridge curve is strictly decreasing, because $v'(u)<0$.  

From \eqref{e:vprime} we also establish that the Beveridge curve is strictly convex, because $v'(u)$ is strictly increasing in $u$. Indeed, the second factor in \eqref{e:vprime} is strictly decreasing in $u$ because $v(u)$ is strictly decreasing in $u$ and $\h>0$. The third factor in \eqref{e:vprime} is also strictly decreasing in $u$ because $\h \in (0,1)$. Since both factors are positive, their product is strictly decreasing in $u$. Given that $-\l/\o<0$, $v'(u)$ is actually strictly increasing in $u$.

\subsection{Proof of proposition~\ref{p:hosios}}

We prove the proposition using the auxiliary function
\begin{equation}
G(r,\t) = \h\t + \frac{r+\l}{q(\t)}.
\label{e:G}\end{equation}
Since $\h>0$, $\l>0$, and $q(\t) = \o \t^{-\h}$ with $\o>0$, $G(r,\t)$ is strictly increasing in $\t$ for any $r\geq 0$.

\paragraph{Characterization of the tightnesses $\t^*$ and $\t^h$} Equation \eqref{e:thetaDmp} shows that the tightness $\t^*$ given by efficiency condition \eqref{e:theta} satisfies
\begin{equation*}
G(0,\t^*) = (1-\h)\frac{1-z}{c}.
\end{equation*}
And \eqref{e:thetaHosios} shows that the tightness $\t^h$ given by the Hosios condition satisfies
\begin{equation*}
G(r,\t^h) = (1-\h)\frac{1-z}{c}.
\end{equation*}
Thus, for any discount rate $r$,
\begin{equation}
 G(0,\t^*) = G(r,\t^h).
\label{e:GEquality}\end{equation}

\paragraph{Zero discount rate} We begin by considering the case $r=0$. Equation \eqref{e:GEquality} implies that $G(0,\t^*) = G(0,\t^h)$, so $\t^*=\t^h$.

\paragraph{Positive discount rate} Next we consider the case $r>0$. We assess the gap between $\t^*$ and $\t^h$ by linearizing the function $G(r,\t)$ around $(0,\t^*)$. Up to a second-order term, we have
\begin{equation}
G(r,\t) = G(0,\t^*)+ \pd{G}{r} \cdot r + \pd{G}{\t}\cdot (\t-\t^*),
\label{e:FLinear}\end{equation}
where the partial derivatives are evaluated at $(0,\t^*)$. We obtain the partial derivatives from~\eqref{e:G}:
\begin{align*}
\pd{G}{r} &= \frac{1}{q(\t^*)} \\
\pd{G}{\t} &= \h +  \frac{\l}{q(\t^*)}\cdot \frac{\h}{\t^*}.
\end{align*}
Using these partial derivatives, we evaluate \eqref{e:FLinear} at $(r,\t^h)$:
\begin{equation*}
G(r,\t^h) = G(0,\t^*)+ \frac{r}{q(\t^*)}  + \h  \bs{\t^* + \frac{\l}{q(\t^*)} }\frac{\t^h-\t^*}{\t^*}.
\end{equation*}
Given that $G(r,\t^h) = G(0,\t^*)$, we find the relative difference between $\t^*$ and $\t^h$:
\begin{equation}
\frac{\t^*-\t^h}{\t^*} = \frac{r}{\h \cdot\bp{f +\l}},
\label{e:difference}\end{equation}
where $f = \t^* q(\t^*) $ is the job-finding rate at $\t^*$. 

\paragraph{Shimer calibration} Last, we quantify the relative difference between $\t^*$ and $\t^h$ using the calibration provided by \ct[table~2]{S05}: $\h = 0.72$, $r = 0.012$ per quarter, $\l = 0.1$ per quarter, and $f = 1.35$ per quarter. Plugging these numbers into \eqref{e:difference}, we find that
\begin{equation*}
\frac{\t^*-\t^h}{\t^*} = \frac{0.012}{0.72 \times \bp{1.35 + 0.1}} = 1.1\%.
\end{equation*}

\section{Beveridge curve in the DMP model}\label{a:distance}

Section~\ref{s:dmp} argues that in the DMP model the labor market is never far from its Beveridge curve. That is, the actual unemployment rate (given by differential equation~\eqref{e:uDotLinear}) is never far from the Beveridgean unemployment rate (the critical point of~\eqref{e:uDotLinear}, given by equation~\eqref{e:ub}). We now illustrate this result using US data for 1951--2019. First, we compute the job-finding rate $f(t)$ and job-separation rate $\l(t)$ that through \eqref{e:uDotLinear} produce the US unemployment rate. Then we compute the corresponding Beveridgean unemployment rate from \eqref{e:ub}: 
\begin{equation}
u^b(t) = \frac{\l(t)}{\l(t)+f(t)}. 
\label{e:ub3}\end{equation}
We confirm that the actual unemployment rate closely tracks the Beveridgean unemployment rate.

\begin{figure}[t]
\includegraphics[scale=\sfig,page=34]{\pdf}
\caption{Quarterly job-finding rate in the United States, 1951--2019}
\note{The job-finding rate is constructed from equations \eqref{e:F} and \eqref{e:f}, as in \ct{S12}. The shaded areas are NBER-dated recessions.}
\label{f:f}\end{figure}

\subsection{Job-finding rate}

To compute the job-finding rate, we follow \ct[pp.~130--133]{S12}. We first construct the monthly job-finding probability:
\begin{equation}
F(t)=1-\frac{u(t+1)-u^{s}(t+1)}{u(t)},
\label{e:F}\end{equation}
where $u(t)$ is the number of unemployed persons in month $t$, and $u^{s}(t)$ is the number of persons who have been unemployed for less than 5 weeks in month $t$ \cp{UEMPLT5,UNEMPLOY}. Assuming that unemployed workers find a job according to a Poisson process with monthly arrival rate $f(t)$, we infer the job-finding rate from the job-finding probability:
\begin{equation}
f(t)=-\ln(1-F(t)).
\label{e:f}\end{equation}
Multiplying this monthly rate by 3, we obtain the quarterly job-finding rate in the United States (figure~\ref{f:f}). Over 1951--2019, the job-finding rate averages $1.76$ per quarter, or $0.587$ per month.

\begin{figure}[t]
\includegraphics[scale=\sfig,page=36]{\pdf}
\caption{Quarterly job-separation rate in the United States, 1951--2019}
\note{The job-separation rate is constructed from \eqref{e:lambda}, as in \ct{S12}. The shaded areas are NBER-dated recessions.}
\label{f:lambda}\end{figure}

\subsection{Job-separation rate}

To compute the job-separation rate, we continue to follow \ct[pp.~130--133]{S12}. The monthly job-separation rate $\l(t)$ is implicitly defined by
\begin{equation}
u(t+1) = \bc{1-e^{-[f(t)+\l(t)]}}\frac{\l(t)}{f(t)+\l(t)} h(t) + e^{-[f(t)+\l(t)]} u(t),
\label{e:lambda}\end{equation}
where $f(t)$ is the monthly job-finding rate (given by \eqref{e:f}), and $h(t)$ and $u(t)$ are the numbers of persons in the labor force and in unemployment \cp{CLF16OV,UNEMPLOY}. Each month $t$, we solve~\eqref{e:lambda} to compute $\l(t)$. Multiplying this monthly rate by 3, we obtain the quarterly job-separation rate in the United States (figure~\ref{f:lambda}). Over 1951--2019, the job-separation rate averages $0.10$ per quarter, or $0.034$ per month.

\begin{figure}[t]
\includegraphics[scale=\sfig,page=39]{\pdf}
\caption{Beveridgean unemployment rate in the United States, 1951--2019}
\note{The Beveridgean unemployment rate is the unemployment rate on the Beveridge curve of the DMP model. It is constructed using equation~\eqref{e:ub3}, the job-finding rate from figure~\ref{f:f}, and the job-separation rate from figure~\ref{f:lambda}. The actual unemployment rate comes from figure~\ref{f:u}; it is displayed as a benchmark. The shaded areas are NBER-dated recessions.}
\label{f:accuracy}\end{figure}

\subsection{Beveridgean unemployment rate}

Finally, we construct the Beveridgean unemployment rate using equation~\eqref{e:ub3}, the job-finding rate from figure~\ref{f:f}, and the job-separation rate from figure~\ref{f:lambda}. The Beveridgean unemployment rate is indistinguishable from the actual unemployment rate (figure~\ref{f:accuracy}). While the maximum absolute distance between the two series is $1.5$ percentage points, the average absolute distance is only $0.2$ percentage point, and the average distance is $0.01$ percentage point.

\section{Endogenous Beveridge elasticity in the DMP model}\label{a:endogeneity}

When we derive the sufficient-statistic formula for the efficient unemployment rate (formula~\eqref{e:u}), we assume that the sufficient statistics do not depend on the unemployment and vacancy rates (assumption~\ref{a:constant}). In the DMP model, however, the Beveridge elasticity depends on the unemployment rate (equation~\eqref{e:epsilonDmp}). But in section~\ref{s:dmp} we argue that the formula should remain accurate because the dependence is weak. Here we confirm this assertion. We calibrate the parameters of the DMP model from US data, 1951--2019. We then compute the efficient unemployment rate in the calibrated DMP model, accounting for the endogeneity of the Beveridge elasticity. We find that the computed efficient unemployment rate is almost identical to the efficient unemployment rate given by formula~\eqref{e:u}.

\subsection{Efficient unemployment rate with endogenous Beveridge elasticity}

In the DMP model, when the endogeneity of the Beveridge elasticity is accounted for, formula~\eqref{e:thetaDmp} gives the efficient tightness $\t^*$:
\begin{equation}
\h\t^* + \frac{\l}{\o} (\t^*)^{\h} = (1-\h) \frac{1-z}{c}.
\label{e:thetaDmp2}\end{equation}
Through the Beveridge curve \eqref{e:ub}, the efficient tightness $\t^*$ and parameters of the model determine the efficient unemployment rate $u^*$:
\begin{equation}
u^* = \frac{(\l/\o)}{(\l/\o)+(\t^*)^{1-\h}}.
\label{e:ub2}\end{equation}

\begin{figure}[t]
\includegraphics[scale=\sfig,page=33]{\pdf}
\caption{Matching elasticity in the United States, 1951--2019}
\note{The matching elasticity is computed using equation~\eqref{e:eta}, the Beveridge elasticity from figure~\ref{f:epsilon}, and the unemployment rate from figure~\ref{f:u}. The shaded areas are NBER-dated recessions.}
\label{f:eta}\end{figure}

\subsection{Application to the United States}

Toward applying formulas \eqref{e:thetaDmp2} and \eqref{e:ub2}, we calibrate the parameters of the DMP model from US data, 1951--2019. 

\paragraph{Social value of nonwork and recruiting cost} As in section~\ref{s:usa}, we set the social value of nonwork to $z = 0.26$ and the recruiting cost to $c = 0.92$. 

\paragraph{Matching elasticity} By inverting equation \eqref{e:epsilonDmp}, we express the matching elasticity $\h$ as a function of the Beveridge elasticity $\e$:
\begin{equation}
\h = \frac{1}{1+\e}\bp{\e-\frac{u}{1-u}}.
\label{e:eta}\end{equation}
We then compute the matching elasticity from the Beveridge elasticity in figure~\ref{f:epsilon} and the unemployment rate in figure~\ref{f:u}. Between 1951 and 2019, the matching elasticity averages $0.44$, and it always remains between $0.39$ and $0.49$ (figure~\ref{f:eta}).

\paragraph{Separation-efficacy ratio} As we assume that unemployment is always on the Beveridge curve, labor flows are balanced, so $\l (1-u) = f(\t) u = \o \t^{1-\h} u$. Therefore, the ratio of the job-separation rate $\l$ to the matching efficacy $\o$ satisfies
\begin{equation*}
\frac{\l}{\o} = \frac{u}{1-u} \cdot \t^{1-\h}.
\end{equation*}
We compute the ratio $\l/\o$ from this relation, the unemployment rate in figure~\ref{f:u}, the tightness in figure~\ref{f:thetaEfficient}, and the matching elasticity in figure~\ref{f:eta}.

\begin{figure}[t]
\includegraphics[scale=\sfig,page=40]{\pdf}
\caption{US efficient unemployment rate with endogenous Beveridge elasticity}
\note{The efficient unemployment rate with endogenous Beveridge elasticity accounts for the endogeneity of the Beveridge elasticity that appears in the DMP model. It is constructed by solving equations~\eqref{e:thetaDmp2} and~\eqref{e:ub2}. The efficient unemployment rate with exogenous Beveridge elasticity comes from figure~\ref{f:uEfficient}; it is displayed as a benchmark. The shaded areas are NBER-dated recessions.}
\label{f:endogenous}\end{figure}

\paragraph{Efficient unemployment rate} Plugging the parameter values into formulas \eqref{e:thetaDmp2} and \eqref{e:ub2}, we compute the efficient unemployment rate in the DMP model (figure~\ref{f:endogenous}). This efficient unemployment rate accounts for the endogeneity of the Beveridge elasticity that arises in the DMP model. Yet it closely tracks the baseline efficient unemployment rate, which takes the Beveridge elasticity as exogenous. The maximum absolute distance between the two series is $0.5$ percentage point, and the average absolute distance is only $0.1$ percentage point.

\section{Hosios condition in the DMP model}\label{a:hosios}

Proposition~\ref{p:hosios} establishes that in the DMP model, the efficient tightness given by the Hosios condition is almost identical to the efficient tightness arising from our Beveridgean approach. Here we simulate a DMP model calibrated to US data, 1951--2019, and we show that the efficient unemployment rates given by the Hosiosian and Beveridgean approaches also are almost identical.

\subsection{Efficient unemployment rate given by the Hosios condition}

In the DMP model, the efficient tightness $\t^h$ given by the Hosios condition satisfies \eqref{e:thetaHosios}:
\begin{equation}
\h\t^h + \frac{\l + r}{\o} (\t^h)^{\h} = (1-\h) \frac{1-z}{c}.
\label{e:thetaHosios2}\end{equation}
Then, the efficient unemployment rate $u^h$ given by the Hosios condition solves differential equation~\eqref{e:uDotLinear}, where the job-finding rate is $f=f(\t^h)$. Accordingly, we compute $u^h(t)$ recursively. We initialize $u^h(1)= u(1)$. We then iterate~\eqref{e:ut}:
\begin{equation}
u^h(t+1) = u^b(\t^h) + [u^h(t) - u^b(\t^h)] e^{-[\l+f(\t^h)]},
\label{e:ut2}\end{equation}
where $f(\t^h) = \o (\t^h)^{1-\h}$ and $u^b(\t^h) =\l/[\l + f(\t^h)]$. 

\begin{figure}[t]
\includegraphics[scale=\sfig,page=35]{\pdf}
\caption{Matching efficacy in the United States, 1951--2019}
\note{The matching efficacy is constructed using equation \eqref{e:omega}, the labor-market tightness from figure~\ref{f:thetaEfficient}, the job-finding rate from figure~\ref{f:f}, and the matching elasticity from figure~\ref{f:eta}. The shaded areas are NBER-dated recessions.}
\label{f:omega}\end{figure}

\subsection{Application to the United States}

To apply formulas \eqref{e:thetaHosios2} and \eqref{e:ut2}, we calibrate the parameters of the DMP model from US data, 1951--2019.

\paragraph{Social value of nonwork, recruiting cost, and matching elasticity} As in appendix~\ref{a:endogeneity}, we set the social value of nonwork to $z = 0.26$ and the recruiting cost to $c = 0.92$, and we take the matching elasticity $\h$ from figure~\ref{f:eta}. 

\paragraph{Job-separation rate} We take the quarterly job-separation rate $\l$ from figure~\ref{f:lambda}. 

\paragraph{Discount rate} As in \ct[table~2]{S05}, we set the quarterly discount rate to $r = 0.012$, which corresponds to an annual discount rate of $5\%$. 

\paragraph{Matching efficacy} With the matching function \eqref{e:matching}, the job-finding rate is $f = \o \t^{1-\h}$, so the matching efficacy satisfies 
\begin{equation}
\o = \frac{f}{\t^{1-\h}}.
\label{e:omega}\end{equation}
We compute the matching efficacy from \eqref{e:omega}, the tightness in figure~\ref{f:thetaEfficient}, the quarterly job-finding rate in figure~\ref{f:f}, and the matching elasticity in figure~\ref{f:eta}; the result is displayed in figure~\ref{f:omega}.

\begin{figure}[t]
\includegraphics[scale=\sfig,page=42]{\pdf}
\caption{Hosiosian efficient unemployment rate in the United States, 1951--2019}
\note{The Hosiosian efficient unemployment rate accounts for the dynamics of unemployment in the DMP model. It is constructed by solving equation~\eqref{e:thetaHosios2} and iterating equation~\eqref{e:ut2}. The Beveridgean efficient unemployment rate comes from figure~\ref{f:endogenous}; it is displayed as a benchmark. The shaded areas are NBER-dated recessions.}
\label{f:hosios}\end{figure}

\paragraph{Efficient unemployment rate} Plugging the parameter values into formulas \eqref{e:thetaHosios2} and \eqref{e:ut2}, we compute the efficient unemployment rate given by the Hosios condition. We find that this Hosiosian efficient unemployment rate is close to the Beveridgean efficient unemployment rate computed in appendix~\ref{a:endogeneity} (figure~\ref{f:hosios}). While the maximum absolute distance between the two series is $1.1$ percentage point, the average absolute distance is only $0.2$ percentage point, and the average distance is $0.03$ percentage point. Moreover, the difference between the two series is not due to differences in the efficient tightnesses; rather, it is due to the conversion of tightness into unemployment.

\section{Fluctuating social value of nonwork in the DMP model}\label{a:fluctuations}

When we compute the US unemployment gap (figure~\ref{f:gap}), we keep the social value of nonwork constant. This choice is justified by the work of \ct{CK16}, who find that the social value of nonwork is acyclical. In some versions of the DMP model, however, the productivities of unemployed and employed workers do not move in tandem over the business cycle, which generates fluctuations in the social value of nonwork. Here we show that such fluctuations have virtually no effect on the efficient unemployment rate.

\subsection{Efficient unemployment rate with fluctuating social value of nonwork}

In the DMP model, the social value of nonwork is constant when the productivity of unemployed workers is proportional to the productivity of labor (equation \eqref{e:zkdmp}). While such proportionality necessarily holds in the long run, it could fail in the short run. In that case, short-run fluctuations in labor productivity are nonneutral: they create fluctuations in the social value of nonwork and in the efficient unemployment rate \cp{S05}. 

To introduce fluctuations of the social value of nonwork in the DMP model, we assume that the productivity of unemployed workers is proportional to the trend of labor productivity, $\bar{p}$, instead of actual labor productivity, $p$. Under this specification, the welfare function~\eqref{e:w} becomes 
\begin{equation}
\Wc(n,u,v) = \bp{p n + \bar{p} z u - p c v} L.
\label{e:w2}\end{equation}
The social value of nonwork becomes
\begin{equation*}
\z = \frac{z}{\hat{p}},
\end{equation*}
where $\hat{p} = p/\bar{p}$ is detrended labor productivity. Formula \eqref{e:u} therefore becomes
\begin{equation}
u^* = \bs{\frac{c \e}{1 - (z/\hat{p})}\cdot\frac{v}{u^{-\e}}}^{1/(1+\e)}.
\label{e:uFluctuations}\end{equation}

\begin{figure}[t]
\subcaptionbox{Labor productivity and trend}{\includegraphics[scale=\sfig,page=37]{\pdf}}\hfill
\subcaptionbox{Detrended labor productivity}{\includegraphics[scale=\sfig,page=38]{\pdf}}
\caption{Labor productivity in the United States, 1951--2019}
\note{A: Labor productivity is the index of real output per worker constructed by the \ct{PRS85006163}. The trend of productivity is produced by a HP filter with smoothing parameter 1600. B: Detrended labor productivity is the labor productivity from panel~A divided by its trend. The shaded areas are NBER-dated recessions.}
\label{f:productivity}\end{figure}

\subsection{Application to the United States}

To apply formula \eqref{e:uFluctuations}, we measure detrended labor productivity and the other statistics in the United States.

\paragraph{Detrended labor productivity} We measure labor productivity ($p$) in the United States from the real output per worker constructed by the \ct{PRS85006163}. We compute the trend of productivity ($\bar{p}$) using a HP filter; since the productivity series has quarterly frequency, we set the filter's smoothing parameter to 1600 \cp{RU02}. We compute detrended labor productivity as $\hat{p} = p / \bar{p}$ (figure~\ref{f:productivity}).

\paragraph{Other statistics} As in section~\ref{s:usa}, we set the average social value of nonwork to $z=0.26$ and the recruiting cost to $c=0.92$, and we take the Beveridge elasticity $\e$ from figure~\ref{f:epsilon}. We also take the vacancy rate $v$ and unemployment rate $u$ from figure~\ref{f:beveridge}. 

\begin{figure}[t]
\includegraphics[scale=\sfig,page=41]{\pdf}
\caption{US efficient unemployment rate with fluctuating social value of nonwork}
\note{The efficient unemployment rate with fluctuating social value of nonwork incorporates the fluctuations of the social value of nonwork that appear in the DMP model when social welfare is given by \eqref{e:w2}. It is constructed using equation \eqref{e:uFluctuations} and the detrended productivity from figure~\ref{f:productivity}. The efficient unemployment rate with constant social value of nonwork comes from figure~\ref{f:uEfficient}; it is displayed as a benchmark. The shaded areas are NBER-dated recessions.}
\label{f:fluctuations}\end{figure}

\paragraph{Efficient unemployment rate} Finally, using formula \eqref{e:uFluctuations}, we compute the efficient unemployment rate in the United States when the social value of nonwork fluctuates over the business cycle (figure~\ref{f:fluctuations}). We find that the efficient unemployment rates with and without fluctuations of the social value of nonwork are indistinguishable. The maximum absolute distance between the two series is only $0.03$ percentage point.

\end{document}